\documentclass[showkeys,preprintnumbers,amsmath,amssymb]{revtex4}

\usepackage{latexsym}
\usepackage{amsmath}
\usepackage{amsfonts} 
\usepackage{epsfig}
\usepackage{graphicx}
\usepackage{dcolumn}
\usepackage{bm}

\newcommand{\Vektor}[1]{\mbox{\boldmath $#1$}}

\newcommand{\pdiff}[2]{\frac{\partial #1}{\partial #2}}

\newcommand{\expval}[1]{\langle #1 \rangle}

\newcommand{\abs}[1]{\left| #1 \right|}

\newcommand{\bea}{\begin{eqnarray}}
\newcommand{\eea}{\end{eqnarray}}
\newcommand{\beann}{\begin{eqnarray*}}
\newcommand{\eeann}{\end{eqnarray*}}

\newcommand{\order}[1]{{\mathcal O}\left( #1 \right)}

\newcommand{\f}[1]{\mbox{\boldmath$#1$}}

\newcommand{\nn}{\nonumber}

\begin{document}

\title{A modelling approach towards Epidermal homoeostasis control}
\author{Gernot Schaller$^*$ and Michael Meyer-Hermann}
\affiliation{
Frankfurt Institute for Advanced Studies (FIAS),
Johann Wolfgang Goethe-Universit\"at,
Max von Laue-Stra{\ss}e 1,
D-60438 Frankfurt am Main,Germany}
\email{schaller@theory.phy.tu-dresden.de}

\date{\today}

\begin{abstract}
In order to grasp the features arising from cellular discreteness and
individuality, in large parts of cell tissue modelling agent-based models
are favoured. 
The subclass of off-lattice models allows for a physical motivation
of the intercellular interaction rules.

We apply an improved version of a previously introduced off-lattice 
agent-based model to the steady-state flow equilibrium of skin.
The dynamics of cells is determined by conservative and drag forces,
supplemented with delta-correlated random forces. 
Cellular adjacency is detected by a weighted Delaunay triangulation.
The cell cycle time of keratinocytes is controlled by a diffusible
substance provided by the dermis.
Its concentration is calculated from a diffusion
equation with time-dependent boundary conditions and varying diffusion
coefficients.
The dynamics of a nutrient is also taken into account by a
reaction-diffusion equation.

It turns out that the analysed control mechanism suffices
to explain several characteristics of epidermal homoeostasis
formation. 
In addition, we examine the question of how {\em in silico} melanoma
with decreased basal adhesion manage to persist within the
steady-state flow-equilibrium of the skin.
Interestingly, even for melanocyte 
cell cycle times being substantially shorter than for keratinocytes, 
tiny stochastic effects can lead to completely different outcomes. 
The results demonstrate that the understanding of initial states of
tumour growth can profit significantly from the application of
off-lattice agent-based models in computer simulations. 

\end{abstract}    


\maketitle


\section{Introduction}\label{Si}

\subsection{Methodology}\label{SSm}

In many applications of mathematical modelling, continuum equations are used
to describe the evolution of discrete biological systems such as tumours,
epithelia, animal populations \cite{murray2002a} etc. Such equations
can be solved comparably efficiently and have helped to understand
many qualitative and quantitative features of tumour growth
\cite{byrne2003b,schaller2006a}. In this modelling approach, the
discrete cell numbers are approximated by continuous functions.
However, the inclusion of discrete effects even in simple
models may lead to qualitatively different results
\cite{bettelheim2001,louzoun2003}.
This problem becomes even more important when modelling the initial
evolution of cancer, which is thought to originate within a single
cell \cite{gatenby2003a}. For systems containing a few cells of a
certain species only, the mathematical foundations of approximating
the species dynamics by a continuous distribution function are rather
shaky, and especially for the complicated interactions between several
cell species including birth and death processes, one may expect a
qualitatively different behaviour to arise from agent-based
approaches. 

Within the class of agent-based (individual-based) models, cellular
tissues are modelled as a set of strongly-interacting discrete
objects. At present, it seems reasonable to consider the cell as the
smallest entity in these models, though this is not stringent (compare
e.~g.~the extended Potts model \cite{graner1992,turner2002}). For agent-based
models, the concept of the cellular automaton
\cite{neumann1966,gardner1970} has proven very useful, since it allows
to describe cells by simple interaction rules. Tumour growth 
for example has widely been modelled with cellular automata
\cite{smolle1998a,dormann2002,ferreira2002}. 
Most of the current agent-based models 
are implemented on a lattice. In models where a single lattice site is
occupied by a single cell, volume-nonconserving events such as
proliferation require far-reaching configuration changes on the
lattice, which in such models comes along with rupture of 
''intercellular bindings'' on a large scale \cite{meineke2001}. 
This leads to the necessity of deriving effective interaction rules,
which can often not be easily related to physical laws. 
Consequently, the avoided lattice artifacts often come at the price of
ending up with parameters that principally cannot be determined by
independent experiments and predictive power of the model is lost.
A physical motivation for cell-cell interactions can be included by
allowing for shape fluctuations. In lattice models, this can be
achieved by increasing the resolution of the lattice (i.~e., by
representing a single cell by a variable number of lattice nodes as
e.~g. in the Potts model \cite{turner2002} or the hyphasma model
\cite{meyer_hermann2004}). 
Likewise, one can also in off-lattice models introduce further degrees
of freedom per cell (such as e.~g. the dynamics of the cell boundary
as in \cite{weliky1990,weliky1991}).
However, especially in the realistic case with three spatial
dimensions, such models with an increased resolution use a
dramatically increased number of dynamic variables and thereby the
necessary calculation time increases (for example, in
\cite{cickovski2005a} about 13000 cells are considered with about 120
lattice nodes per cell in a three-dimensional setup). As an advantage however,
physically-motivated interactions can be used to calculate the
cellular dynamics.  
Off-lattice centre-based models \cite{drasdo1995,dallon2004} allow for
physical interactions to be included without the disadvantage of the
large number of dynamic variables. As a drawback, they assume an
intrinsic cellular equilibrium shape (e.~g. spherical or ellipsoidal)
and treat all deviations from this form as small perturbations. Thus,
they represent a compromise between speed of numerical calculation and
model accuracy. 

We have previously applied an off-lattice centre-based model to the
qualitatively simple problem of multicellular tumour spheroid growth
\cite{schaller2005a}. In comparison to an analogous continuum approach
\cite{schaller2006a} it turned out that agent-based models have
greater potential to describe specific biological systems
realistically. Keeping in mind that continuum models also arise from
averaging (thereby deleting information), this is not a big
surprise. However, it must be said that the degree of model complexity
should always be limited by the experimental signature and for many
experimental signatures continuum models will suffice. 

In this article, we set up an off-lattice centre-based model for the
epidermis that is intrinsically consistent and is also based on
physical interactions as far as possible. This has the advantage that
the model can be falsified. 


\subsection{The Epidermis}\label{SSe}

The epidermis is a stratified squamous epithelial tissue.
It does not contain separate blood vessels and is therefore
dependent on diffusion of nutrients from the dermis situated
below. The epidermis can be divided into several layers
\cite{montagna1992}: 

The innermost {\em stratum germinativum} or {\em stratum basale}
(basal layer) is a monolayer, in  
which most cell divisions occur. It is separated from the dermis below
by a basal membrane.
A fraction of the cells created there by cell division travels upwards
into the {\em stratum spinosum}.
Within this layer, the process of cornification 
begins: The cytoplasm looses water and is filled with keratin
filaments.
Within the {\em stratum granulosum}, cells die off
and their shape flattens. This special pathway of cell death is also
called anoikis.
Completely cornified cells mark the {\em stratum corneum}, which is
clearly distinguishable from the layers below. This layer does not
contain viable cells and constitutes an efficient barrier for water
and many of its solutes. The thickness of this layer varies strongly
for different regions of the skin \cite{montagna1992}.
In the upper part of this layer, the cellular material detaches by
dissolving intercellular contacts.

Within this article, only three layers will
be distinguished: The term {\em stratum medium} will be used as a
combination for all layers not belonging to the {\em stratum germinativum} or
the {\em stratum corneum}.

The cell types encountered in the epidermis are keratinocytes,
melanocytes, Langerhans cells, and Merkel cells. Of these, the
dominant fraction is constituted by the keratinocytes with roughly
75000 cells per square mm \cite{hoath2003,bauer2001}.

Keratinocytes are produced in the {\em stratum germinativum} by cell
division. In order to maintain epidermal homoeostasis, in average one
of the two daughter cells must leave the basal layer and
travel upwards. The keratinocyte remaining in the basal layer will be
termed stem cell in this article. 
The cell travelling upwards transforms into a fully-differentiated
keratinocyte -- possibly undergoing transit amplifying proliferations
\cite{barthel2000} -- and reaches the surface after about 12 to 14
days. During this passage, the keratinocytes follow the process
of cornification described above.

Melanocytes are dendritic cells that are distributed within the basal
layer, and their density is approximately 2000 cells per square mm
\cite{montagna1992,moncrieff2001,hoath2003}.
They adhere to the basal membrane via hemi-desmosomes. 
Their normal function is to provide keratinocytes in the skin with
melanin. Tumours arising from cancerous melanocytes are called
melanoma. Since most cancerous melanocytes still produce melanin, such
tumours have a characteristic black colour.

The Langerhans cells are dendritic cells of the immune system and it
is believed that Merkel cells play a role in sensation.
Since neither effects of the immune system nor the mechanisms of
sensation will be studied here, the latter two cell types will not be
contained in the {\em in silico} representation and will not be
discussed further. 

The diffusional properties of the skin have important implications on
medication applied to this tissue. With an observed strong 
increase of the manifestation of melanoma \cite{moncrieff2001},
studies of melanoma development are of huge importance. Especially
their early diagnosis bears the potential to improve the chances of
recovery. Within this article, some basic questions related to the
initial stages of melanoma growth will be addressed.


\section{The model}\label{Smm}

In view of the complicated matter in reality, any model will
inevitably simplify the system by neglecting properties that we
consider to be less important. Starting from this perspective, the
following approximation may seem reasonable:

Within the model, cells are represented as adhesively and elastically
interacting (compressible) spheres of time-dependent radii. Processes
such as cell proliferation and cell death correspond to  insertion or
removal of spheres to the system, respectively. The cell
growth determines the time-dependent cell radii. Accordingly, cells
consume nutrients and their presence also changes the diffusive properties
within the tissue. In analogy to the cell cycle, the model cells can assume different
internal states.

Note that the simplifying assumption of spherical cell shapes
facilitates the use of adjacency detection algorithms such as the
Delaunay triangulation \cite{schaller2004}.
The basic specifics and consequences of this paradigm will be examined
in the following, whereas the detailed technical discussion can be
found in the appendix.


\subsection{Continuous model variables}\label{SScmv}

The problem of describing the dynamics of cell-cell contacts is
currently not solved. In this article we have chosen the
Johnson-Kendall-Roberts (JKR) model \cite{johnson1958,johnson1971} --
supplemented with viscous shear and normal forces and delta-correlated
random forces. 
The JKR model includes elastic and adhesive normal forces and has been
experimentally verified for soft materials such as rubber.
The simultaneous treatment of elastic and adhesive properties goes
beyond some previous modelling approaches
\cite{kreft2004a,picioreanu2004a,schaller2005a}.
Cell movement within solution (and even more in tissue) can be
regarded as highly overdamped, which is why viscous effects cannot be
neglected (in fact, they are known to be dominant). 
A detailed discussion of these interactions can be
found in appendix \ref{ASjkr} for the JKR-model, appendix \ref{ASrf} for
the random forces, and appendix \ref{ASeom} for the arising equations
of motion including dampening force contributions.

Under the assumption of dominant friction and neglecting the
contribution of torque to dampening, the equations of motion
for $N$ cells can be written as a large $3N \times 3N$
matrix $A(t)$ acting on the $3N$-dimensional cell position 
vector 
\mbox{$\Vektor{x}=\left[x_1(t), y_1(t), z_1(t), \ldots, x_N(t), y_N(t), z_N(t)\right]$}
\bea\label{Eeqnone}
A(t) \dot{\Vektor{x}} = \Vektor{b}\,,
\eea
where the vector $\Vektor{b}(t)$ on the right hand side includes all
non-viscous forces and the matrix $A(t)$ contains the dampening
contributions, see also the example in the appendix \ref{ASns}.
Since there is no analytic form of these quantities, equation (\ref{Eeqnone}) has
to be solved numerically.

As the friction matrix $A(t)$ has neither a diagonal (non-isotropic
friction) nor another simple structure, its inverse cannot be easily
calculated. The fact that only cells in contact can contribute to
friction however leads to a sparse population of the dampening matrix. In
addition, since more than one spatial dimension is considered, the
dampening matrix is not even block diagonal. 
Therefore, the inverse of such a sparse matrix is not necessarily
sparse as well and for the systems considered in this paper, the
inverse matrix of $A(t)$ would even not fit into main memory of normal
PC's. Therefore, we have used an
iterative procedure -- the method of conjugate gradients, see appendix
\ref{ASns} -- that does not necessitate the explicit calculation of
$A^{-1}(t)$ to solve above equation for the position vector
$\Vektor{x}(t)$.  

In addition, one has to solve the problem of the dynamics of
diffusible signals such as e.~g. nutrients. If processes such as
convection and flow transport are comparably small, these are well
described by reaction-diffusion equations. Such types of equations
arise naturally from the continuum equation, if one assumes that the
flow is always proportional to the concentration gradient and tends
to even out all gradients.
The discretisation of reaction-diffusion equations on a regular
lattice leads to a large linear systems as above. However, in this
case the matrix possesses more symmetry properties and
well-adapted algorithms exist for the solution, see appendix
\ref{ASns}.
Note that an alternative approach circumventing a grid discretization 
would be to use the method of Green functions following \cite{newman2004}.

Within the JKR model, adhesion is described by an adhesion energy
density parameter $\sigma$. Assuming spatially uniform receptor
($C_i^{\rm rec}$) and ligand ($C_i^{\rm lig}$) densities on the cell
membranes of cells $i$ and $j$, this parameter can be expressed as
\mbox{$\sigma_{ij} = \frac{\sigma^{\rm max}}{2} 
\left[C_i^{\rm rec} C_j^{\rm lig} + C_i^{\rm lig} C_j^{\rm
rec}\right]$}, where $\sigma^{\rm max}$ determines the maximum
adhesion energy density (model parameter). 
A measure for the total cellular binding strength can then be derived from
the sum of all binding energies with the next neighbours
\bea
\Sigma_i(t) = \sum_{j \in {\cal NN}(i)} \sigma_{ij}(t) A_{ij}(t).
\eea
Assuming that cells with low binding are shed off the skin
surface, we remove necrotic and cornified cells from the simulation as
soon as their binding strength falls below a critical value 
$\Sigma_i \le \Sigma^{\rm min}$ (model parameter).
Note that this choice also has the consequence that non-viable cells
without contact to other cells are also removed from the simulation. These
cells do not have anchorage and would be shed off in the
realistic epidermis.
In the case of cornified and necrotic cells we have assumed an exponential 
decay of receptor and ligand molecules with a given rate $\alpha$, i.~e.,  
\bea\label{Erecligrate}
\dot{C}_i^{\rm rec/lig} = -\alpha C_i^{\rm rec/lig}\,.
\eea
In the JKR model [compare equation (\ref{Ejkr_model})], the resulting decreased
cell-cell adhesion would -- with unchanged elastic parameters -- lead
to a perturbation of the equilibrium distance. Therefore, we have
chosen to adapt the elastic cell modulus simultaneously. To maintain
for similar cells a constant equilibrium distance, this implies a
decreasing elastic modulus according to 
\bea
\dot{E}_i = - 2\alpha E_i\,.
\eea

Although the loss of receptors and ligands as well as decreasing cell
elasticity may be reasonable assumptions for cornified and necrotic
cells, the overall time course may be quite different in reality. 
We have tried other forms of necrotic cell removal. For example, one
could think of removing non-viable cells randomly at a constant rate
as was done in \cite{schaller2005a}. This possibility however did
significantly disturb the layered structure of the {\em stratum
corneum}. Holes in this protective layer in turn did lead to sudden
loss of the proliferation-moderating soluble substance in the epidermis and
thereby to irregular proliferative behaviour and considerable
oscillations in epidermal thickness. The same problem occurred when assuming a
(Gaussian-distributed) cell-specific time after which non-viable
cells were removed from the simulation.

Thus, the assumptions regarding continuous model variables can be
summarised as follows:

\begin{itemize}
\item cell shape : spherical with slight perturbations
\item cell-cell interactions : elastic, adhesive, and viscous non-isotropic dampening
\item approximations : dominating friction
      and neglect of the back-reaction of angular velocities on the
      kinetics 
\item necrotic and cornified cells : loss of receptor and ligands
      leading to a reduced intercellular adhesion 
\end{itemize}


\subsection{Discrete Model Variables}\label{SScc}

\begin{figure}[htb]
\begin{tabular}{cc}
\begin{minipage}{0.5\linewidth}
\includegraphics[height=6cm]{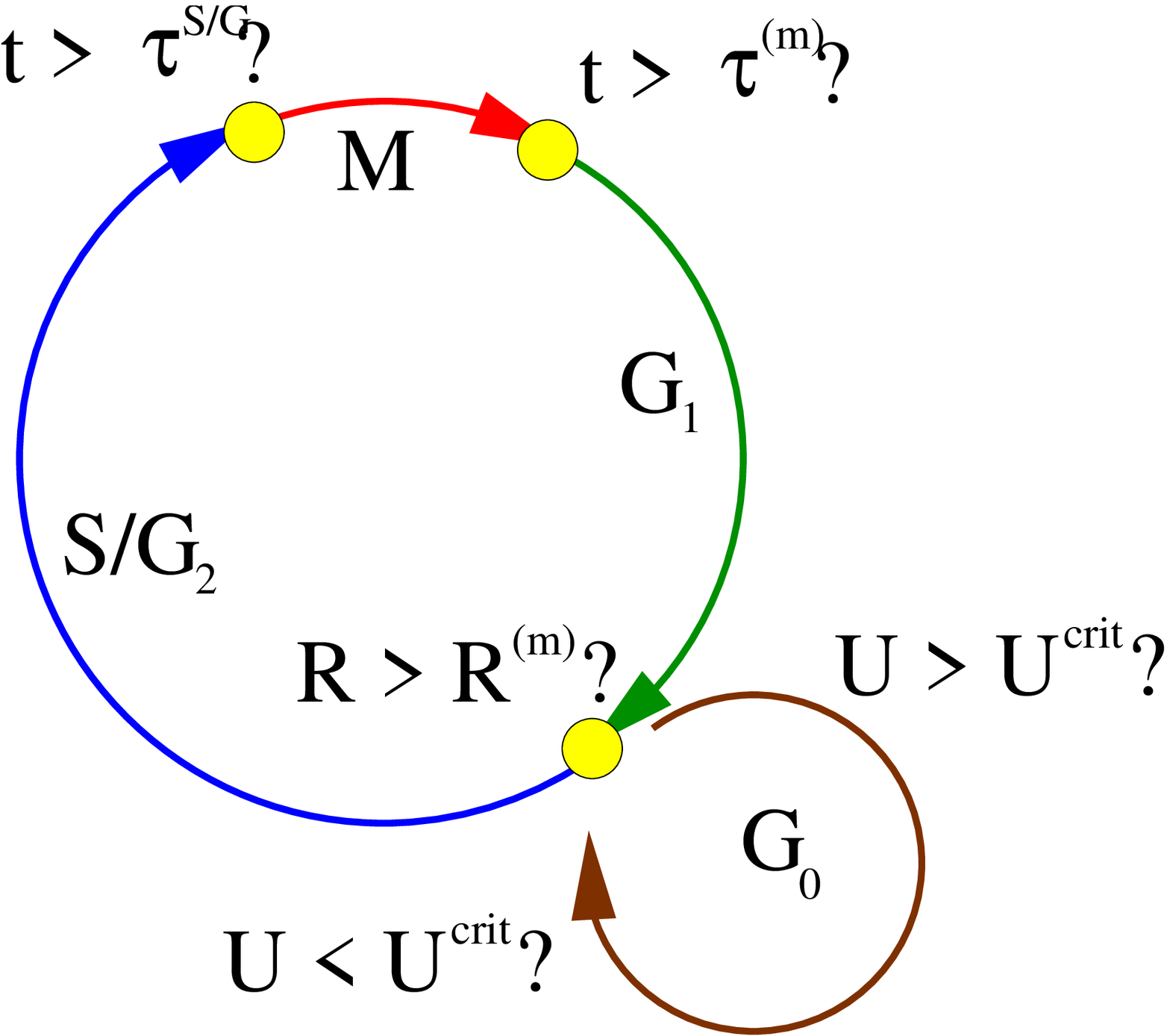}
\end{minipage}
&
\begin{minipage}{0.5\linewidth}
\includegraphics[height=6cm]{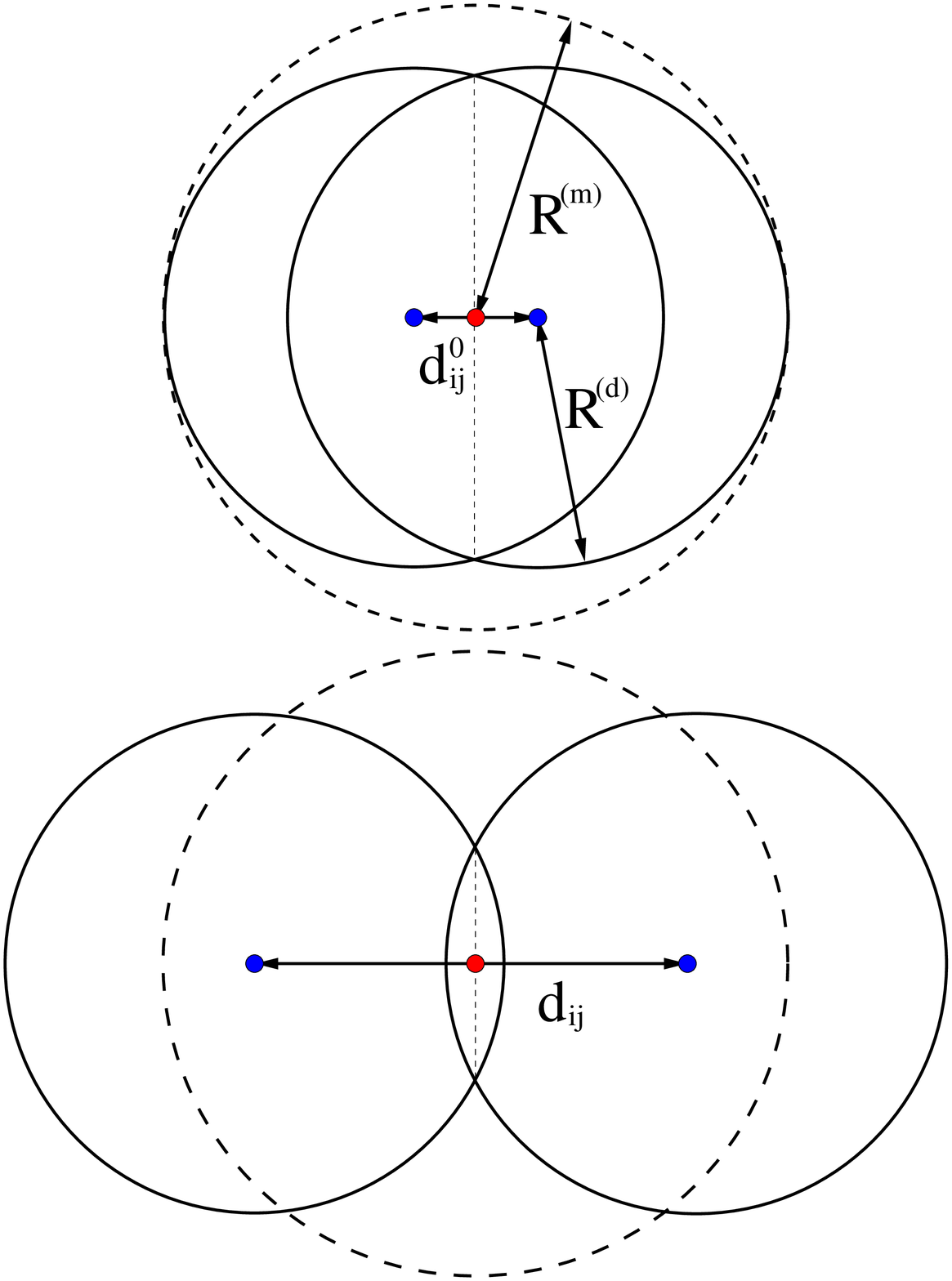}
\end{minipage}
\end{tabular}
\caption{\label{Fcyclesketch}
{\bf Left}: Model realization of the cell cycle.
During cell division, cells reside in the
M-phase for $\tau^{\rm (m)}$. Afterwards, the cell volume increases at a
constant rate in the $\rm G_1$-phase, until the pre-mitotic radius
$R^{\rm (m)}$ has been reached. At the end of the $\rm G_1$-phase, the
cell can either directly continue with the cell cycle or it can prolong the
cycle time by entering the $\rm G_0$-phase, if the local concentration
of the regulative substance (e.~g.~mobile water)
exceeds a threshold $U^{\rm crit}$. The $\rm G_0$-phase
is left after an individual time drawn from a Gaussian distribution or
if the local concentration of the regulating signal falls below the
critical threshold. 
Depending on the local nutrient concentration, cells can
enter necrosis at any time in the cell cycle (not shown). In addition,
keratinocytes enter anoikis in the fourth cell generation after
completion of the $\rm S/G_2$-phase.
{\bf Top right}: Displayed is the configuration right at cell division
at the beginning of the M-phase. The radii of the daughter cells
(solid circles) are reduced to ensure conservation of the target volume.
{\bf Bottom right}: At the end of the M-phase, the daughter cells
have relaxed. Due to interaction with neighbouring cells (not shown),
the direction of mitosis may generally change during M-phase. An
adaptive timestep in the numerical solution ensures that contact is maintained. 
}
\end{figure}

Without representations of internal cellular states, the
model would merely calculate the mechanical interaction between a
number of adhesively  and elastically interacting deformable
spheres. However, it is well known that the different states in the
cell cycle yield a different cell behaviour. This should also be
reflected in the model.
Extending a previous agent-based modelling approach
\cite{schaller2005a} to the necessities of the epidermis, we 
distinguish between the following internal states in the model: 
M-phase, $\rm G_1$-phase, $\rm S/G_2$-phase,
$\rm G_0$-phase, necrotic, cornified. The first four states
are illustrated in figure~\ref{Fcyclesketch} left panel.
During $\rm G_1$-phase, the cell volume grows at a constant rate
$r_{\rm V}$,
i.~e., the radius increases according to 
\mbox{$\dot{R} = \left(4 \pi R^2\right)^{-1} r_{\rm V}$}, 
until the cell reaches its final mitotic radius
$R^{\rm (m)}$. The volume growth rate $r_{\rm V}$ is deduced 
from the minimum observed cycle time 
$\tau^{\rm min}$ and the durations of the $\rm S/G_2$-phase and the M-phase.
Afterwards, no further cell growth is performed.

At the end of the $\rm G_1$-phase, a checkpointing mechanism is performed.
At this checkpoint, the cell can either enter the $\rm G_0$-phase
or the $\rm S/G_2$-phase: In \cite{potten2000} it has been suspected that a 
diffusible substance produced at the basal layer might moderate the cellular 
proliferation. It is well known that the {\em stratum corneum} constitutes an 
effective barrier against the loss of water and its solutes as well as other
substances \cite{barthel2000,bashir2001,kasting2003}. Its removal leads
to a proliferative response. Hence, we use this correlation to
establish a causal connection as a hypothesis. As the simplest
model assumption, we will here assume the distribution of 
{\em mobile water}
to influence cell proliferation. By mobile water we mean the
fraction of the water content of the tissue that is not bound in
intra- or extracellular cavities. Note that evidently the
concentration of this water fraction can change in time and space. 
If the local mobile water concentration is below a
critical value (model parameter), the cell will directly continue the
cell cycle, whereas in the other case the cell cycle will be prolonged
by the cell switching into the $\rm G_0$-phase.
Cells leave the $\rm G_0$-phase to enter the $\rm S/G_2$-phase if either the local water
concentration falls below the threshold or after an individual maximum
time has passed (drawn from a random number generator, see
subsection~\ref{SSstochastic}). 
Note that the assumption of a different moderating diffusible signal
would not significantly change the model as long as it is not created
or consumed by the cells in the epidermis themselves.

Within this article the S-phase and $\rm G_2$-phase are not
distinguished, their inclusion would be a mere technical aspect.
After leaving the $\rm S/G_2$-phase (see
subsection~\ref{SSstochastic}), the cells deterministically enter
mitosis. Keratinocytes and healthy melanocytes underly an exception at
this point: After the fourth cell generation, keratinocytes cornify
(enter anoikis) \cite{meineke2001,potten2000,barthel2000}. Healthy
melanocytes simply remain at the end of $\rm S/G_2$-phase.

Within the model, the difference between the $\rm S/G_2$-phase
and the $\rm G_0$-phase is that the duration of the first is
determined by an individual duration that can be
derived from experiments, whereas the duration of the latter is also
controlled by the spatio-temporal evolution of the concentration of
the moderating substance (in this case, mobile water).

One should be aware that our classification of internal cellular
states may not directly correspond to the realistic biological
system. However, the only net effect of the existence of the $\rm G_0$-phase
is the prolongation of the cell cycle time: Cells in 
$\rm G_0$-phase can serve as a 
reservoir of cells ready to start proliferating as soon as the local
water concentration falls below a critical threshold.
A different terminology or a placement of the
$\rm G_0$-phase after or within the $\rm S/G_2$-phase would therefore
not significantly change the model.

At the beginning of the mitotic phase -- which lasts for about half an
hour for most cell types -- a mother cell is replaced by two daughter cells.
The radii of the daughter cells are decreased $R^{\rm (d)} = R^{\rm
(m)} 2^{-1/3}$ to ensure conservation of the target volume during
M-phase. In addition, they are placed at distance 
\mbox{$d_{ij}^0 = 2 R^{\rm (m)} (1-2^{-1/3})$} to ensure that
in this first discontinuous step the daughter cells do not leave the region
occupied by the mother cell, see figure~\ref{Fcyclesketch} right panels.
In most cases, the daughter cells have the same cell type as the
mother cell. The only exception is given by the keratinocyte stem cells which
divide asymmetrically: By model assumption the upper daughter
cell differentiates to a keratinocyte. 
The new cells are subject to their initially dominating repulsive forces
(\ref{Ejkr_model}). Note that an adaptive timestep derived from a
maximum spatial stepsize ensures that the mitotic partners do not
loose contact. 
Afterwards, the daughter cells enter the $\rm G_1$-phase thus closing
the cell cycle.

Viable cells can enter necrosis at any time in the cell cycle as soon
as the nutrient concentration at the cellular position falls below a
cell-type specific critical threshold (model parameter).
As the dominant pathway to cell death, keratinocytes in contrast enter
anoikis after completing $\rm S/G_2$-phase in the fourth generation.
Naturally, necrotic or cornified cells do not consume any nutrients.

The corresponding assumptions on discrete model variables can be
summarised as follows

\begin{itemize}

\item cell proliferation
\begin{itemize}
\item cellular states: M-phase, $\rm G_1$-phase, $\rm S/G_2$-phase,
$\rm G_0$-phase, necrotic, cornified
\item local mobile water concentration can prolong the duration of
      $G_0$ state for keratinocytes
\item conservation of target volume during M-phase
\item for stem cells: upper cell differentiates to a keratinocyte,
      lower cell remains a stem cell
\item keratinocytes can undergo a maximum of four transit
      proliferations, whereas stem cells divide {\em ad infinitum}
\item healthy melanocytes do not proliferate, whereas malignant
      melanocytes can divide {\em ad infinitum}
\end{itemize}

\item cell growth: growth of cell volume at constant rate during 
	$\rm G_1$

\item cell death
\begin{itemize}
\item low local nutrient concentration induces necrosis
\item keratinocytes undergo cornification after fourth generation
\item cornified cells without contact to others are removed from the
      simulation immediately
\end{itemize}

\end{itemize}


\subsection{Stochastic Elements}\label{SSstochastic}

It is an empirical fact that processes in biological systems
underly significant stochastic deviations: For example, biofilm cell
populations starting from a single cell desynchronise proliferation
after about five generations \cite{kreft1998}. Such a behaviour can not
be explained by processes such as contact inhibition or nutrient
depletion, as these are not active for small systems with only $2^5$
cells.

In the model, this is represented by stochastic elements that can be
derived from a pseudo-random number generator \cite{matpack_manual}. 
The involved stochastic elements are the delta-correlated
random forces (see appendix \ref{ASrf}) acting on every cell, the
initial direction of the displacement vector at mitosis, and the
durations of some cell cycle phases such as the M-phase, the $\rm
S/G_2$-phase, and the $\rm G_0$-phase. 

As was done in previous models for biofilms \cite{kreft2001a,picioreanu2004a},
the initial direction of mitosis is determined from a random
distribution, which is unifom on the unit sphere. This is the simplest
modelling assumption that did not induce artifacts. In addition, it
should be noted that during M-phase configuration changes are still
possible due to interactions with the neighbouring cells.

In order to yield a sufficiently fast desynchronisation of the cell
cycles, the individual duration times for the M-phase and the $\rm
S/G_2$-phase as well as the maximum duration time for the $\rm
G_0$-phase are drawn from a normally-distributed random number
generator \cite{matpack_manual} with a given mean and width.
Without these stochastic elements, the model exhibits artificial
oscillations around a steady state even in later stages. 
Technically, the duration of each phase is determined at the beginning
of the phase. 
Naturally, the parameters on the random number generators can be set
individually for every cell type.


\subsection{Computer Platform}

The computer code was written following the paradigm of
object-oriented programming in $\rm C^{++}$ and was
compiled with the GNU compiler {\tt gcc} version 3.3.
The code was executed on an AMD Athlon(tm) MP Processor 1800+ with 1
GByte of RAM on a Linux platform.


\subsection{Simulation setup}\label{Sss}

As the computational domain, a rectangular box of dimensions 
\mbox{$\rm 200 \;\mu m \times  200 \;\mu m \times  400 \;\mu m$} has been
considered. 
Since epidermal tissue is anisotropic, the boundary conditions have to
be chosen non-homogeneous as well.
Note that the cellular kinetics is described with a system of ordinary
differential equations (\ref{Eeom}). Therefore, the term ``boundary
condition'' refers to the special interactions of cells with the
boundary of the computational domain.
It is known that a realistic epidermis exhibits a ruffled basal layer
\cite{montagna1992}. However, in order to treat the microenvironment of
epidermal tissue as simple as possible, the basal layer has been
implemented here as a static planar boundary at the bottom with normal
vector $\Vektor{e}_z$.  
With using the JKR model (\ref{Ejkr_model}), the interaction with
such a planar boundary can be well implemented by assuming contact
with a cell of infinitely large radius. 
Specifically, the $z$-boundary has been assumed to be of
infinite elasticity $E_{\rm bound}=\infty$. Since the inverse elastic
moduli enter additively in the JKR model in equation 
(\ref{Ejkr_details}), this choice does not sensitively change the
global model behaviour but merely shifts the 
equilibrium distance between basal membrane and bottom cell layer. The
corresponding adhesive anchorage in the basal layer has been made
dependent on the cell type (see the discussion below). 
In order to minimise the boundary effects in $x$ and $y$ direction, periodic
boundary conditions could be used for the cell cell interaction. This 
however would necessitate a rather tedious mirroring of cells close to
the boundary. In addition, one would have to use periodic boundary
conditions on the associated reaction-diffusion equations as well to
avoid additional artifacts.
Therefore, here a different (mirror cell) approach has been chosen:
Every cell in contact with a $x$ or $y$ boundary is assumed to be in
contact with a cell of the same type, size, receptor and ligand
equipment, etc. In short, it interacts with a virtual mirror copy of itself,
where the contact area is situated within the boundary plane. 
In upper $z$-direction there are no boundary conditions on the cells
-- recall that necrotic or cornified cells are removed
eventually. In comparison to a static boundary this procedure also has
the additional advantage that drag force artifacts are reduced.

The boundary conditions on the cells have their counterpart in the
reaction-diffusion equations for the mobile water concentration and
the nutrients:
The concentrations at the lower $z$-boundary have been fixed to the
maximum value (Dirichlet boundary conditions), and above the cell
layers (dynamic thickness, a {\em stratum corneum} need not always
exist during the simulations) both 
concentrations are fixed to 0. Technically, this has been implemented
by setting the concentrations to vanish at all grid volume elements
not containing any cells: The resolution of the reaction-diffusion
grid was low enough to prevent the emergence of artificial sink terms
in intercellular cavities throughout all simulations
(such problems could -- in principle -- also be avoided completely 
by using Green functions \cite{newman2004}).
At the $x$ and $y$
boundaries, no-flux von Neumann boundary conditions have been used,
i.~e., $\partial_x u = 0$ and $\partial_y u = 0$. Note that this
is equivalent to the corresponding boundary conditions on the cells: The
boundary is impenetrable for both cells and nutrients. Thus, for an in
$x$ and $y$ directions homogeneous cell distribution, the problem
would effectively reduce to a one-dimensional one. 

The initial conditions have been determined as follows:
A monolayer of keratinocyte stem cells was distributed on the basal
membrane. Afterwards, the position of the cells in the
cell cycle has been randomised uniformly to avoid initial
artifacts. This configuration could for example mimic a severely
perturbed epidermis, where suddenly not only the {\em stratum corneum}
but also the  {\em stratum medium} was removed. Consequently, a strong
proliferative response should be expected. 

After establishment of a steady-state flow equilibrium, different
perturbations have been performed. These will be discussed in the next
section.


\section{Results}\label{Sr}


\subsection{Flow equilibrium}\label{SSfe}

\begin{figure}[htb]
\begin{tabular}{cc}
\begin{minipage}{0.5\linewidth}
\includegraphics[height=7cm]{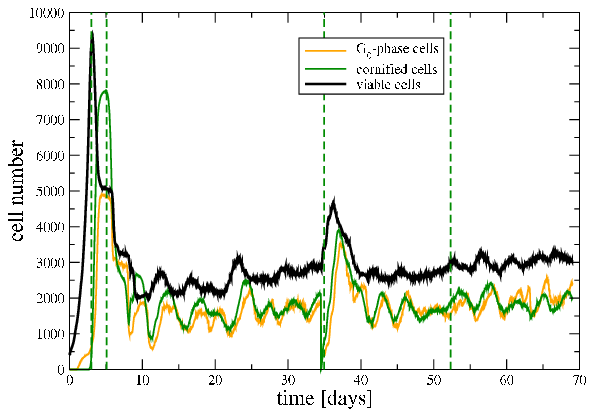}
\end{minipage}
&
\begin{minipage}{0.5\linewidth}
\includegraphics[height=7cm]{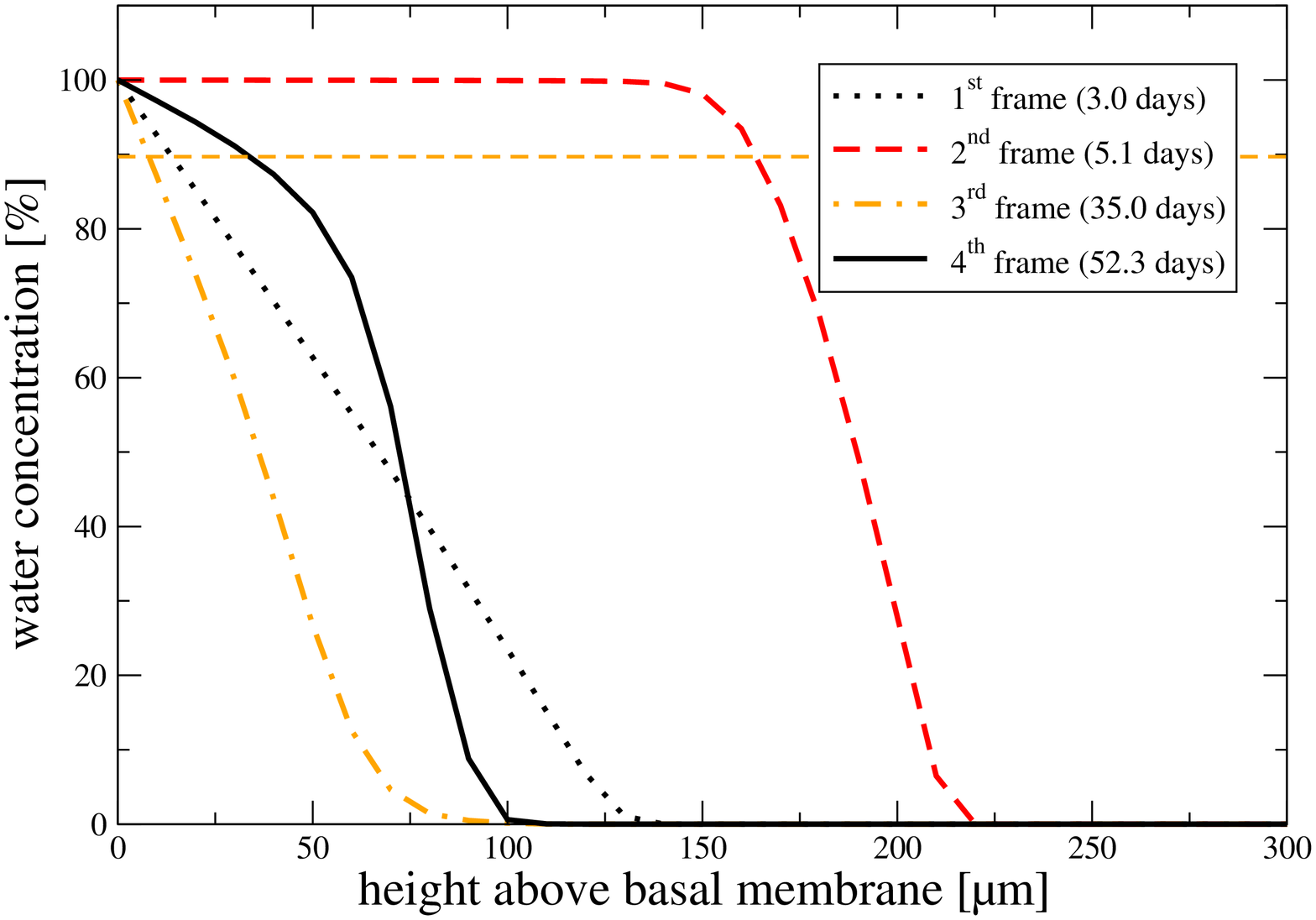}
\end{minipage}
\end{tabular}
\includegraphics[width=14cm]{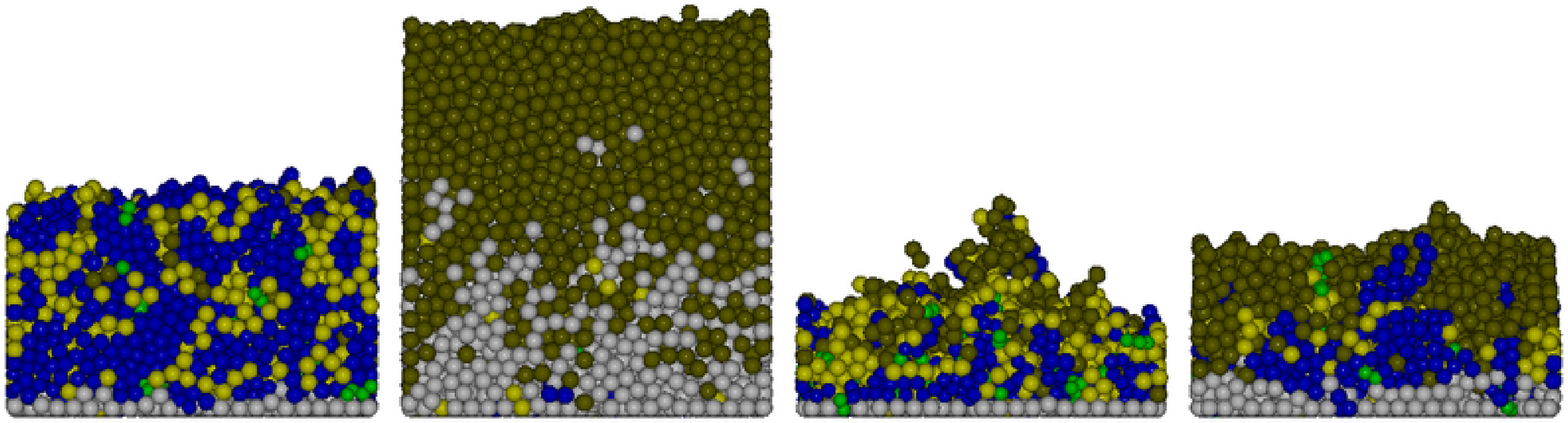}
\caption{\label{Ftape_stripping}
Emergence of an epidermal steady-state flow equilibrium.
{\bf Top Left}: 
Distribution of keratinocytes in the cell cycle. Cell numbers refer to
a ground surface of $\rm 200 \times 200\;\mu m^2$. Vertical dashed lines
indicate times where cross-sections of the keratinocyte distribution
(bottom row) and the water concentration (top right) have been extracted.
{\bf Top Right}: Shown is the (in $x$ and $y$ averaged) dependence of
the concentration of mobile water on the distance from the basal layer at
different times. The maximum value of 100 \% corresponds to the mobile
water concentration at the basal membrane and the horizontal dashed line
denotes the critical mobile water concentration $U_{\rm H_20}^{\rm
crit}=90$ \%.
{\bf Bottom row}: Shown is a cross-section in the $x-z$-plane 
of the {\em in silico}
epidermis. Cells in $\rm G_0$-phase are marked in light grey,
cornified keratinocytes are marked in grey, and darker shades of grey
denote the different phases of the cell cycle. The first frame
corresponds to an excited state already, whereas the last frame
displays the flow equilibrium distribution predicting an approximate
thickness of $120 \;\mu m$ above the basal membrane.
}
\end{figure}

Our first question was whether the proposed control mechanism of the
water-concentration-induced prolongation of the cell cycle time could
actually produce the macroscopically observed 
flow equilibrium of skin. In particular, we asked whether
\begin{itemize}
\item
a steady-state flow equilibrium is established, and
\item 
whether this equilibrium is stable against perturbations such as
complete removal of the {\em stratum corneum} that is performed for
example in tape-stripping experiments \cite{barthel2000}.
\end{itemize}
These questions can be interpreted as a sanity check of the model
assumptions and it turns out that both have an affirmative answer (see
figure~\ref{Ftape_stripping}).  
Starting from a monolayer of cells, the local water concentration
above the basal membrane is quite low such that no cell enters cell
cycle prolongation. The net effect is an 
initial exponential growth phase (top left panel). After four generations,
cornification of the first keratinocytes begins, followed by 
the formation of a {\em stratum corneum} with a considerably decreased
diffusion coefficient for water. This in turn leads to an increased
water concentration and thereby a greater fraction of cells residing
in $\rm G_0$-phase: The initial exponential growth slows down and then 
the cell number decreases, since the cornified cells shed off the
skins surface. Afterwards, the dynamics equilibrates. After 35 days, a
tape-stripping experiment has been performed: All cornified cells are
suddenly removed from the simulation. This leads again to a
proliferative response. However, since this time the cornified layer
quickly re-establishes due to reservoir cells in $\rm G_0$-phase, the
proliferative response is considerably smaller than
initially. Note that the dominant contribution to the rapid formation
of the cornified layer in the model results from the fraction of $\rm
G_0$-keratinocytes that have already reached their fourth generation.
Interestingly, the oscillations around the equilibrium
value are remarkably strong. The number of cells displays a
slight (but decelerating) upward tendency, but 15 days after the
disturbance (last vertical line), saturation is nearly reached.
The final cell numbers correspond well to observed densities of
keratinocytes (75000 cells per square mm skin at the breast \cite{bauer2001}).
In the top right panel of figure~\ref{Ftape_stripping} it is
demonstrated that with an intact {\em stratum corneum} (second and
last frame), the water concentration is large in the lower layers of
the epidermis and then falls rapidly.
In figure~\ref{Ftape_stripping} bottom
row it becomes visible in the latest frame that the cornified layer 
exhibits a small hole (cells in dark grey). Due to a considerable loss
of water, this causes many distant keratinocytes to leave their cell
cycle arrest (cells in light grey changing to cells in dark grey) and
thereby leads to a perturbation of the equilibrium. 

In \cite{potten2000} the authors had hypothesised a diffusible
substance that moderates cellular proliferation times. The present
model does not contradict the hypothesis that this substance could
simply be the moisture of the epidermis but other diffusible
substances would presumably lead to equivalent model
behaviour. Therefore, a confirmation/falsification of this model
hypothesis would require more data on the candidate substances
(diffusion coefficients, reaction rates, concentrations) and the
associated processes.


\subsection{Melanocyte anchorage}\label{SSma}

\begin{figure}[htb]
\begin{tabular}{cc}
\begin{minipage}{0.5\linewidth}
\includegraphics[height=6cm]{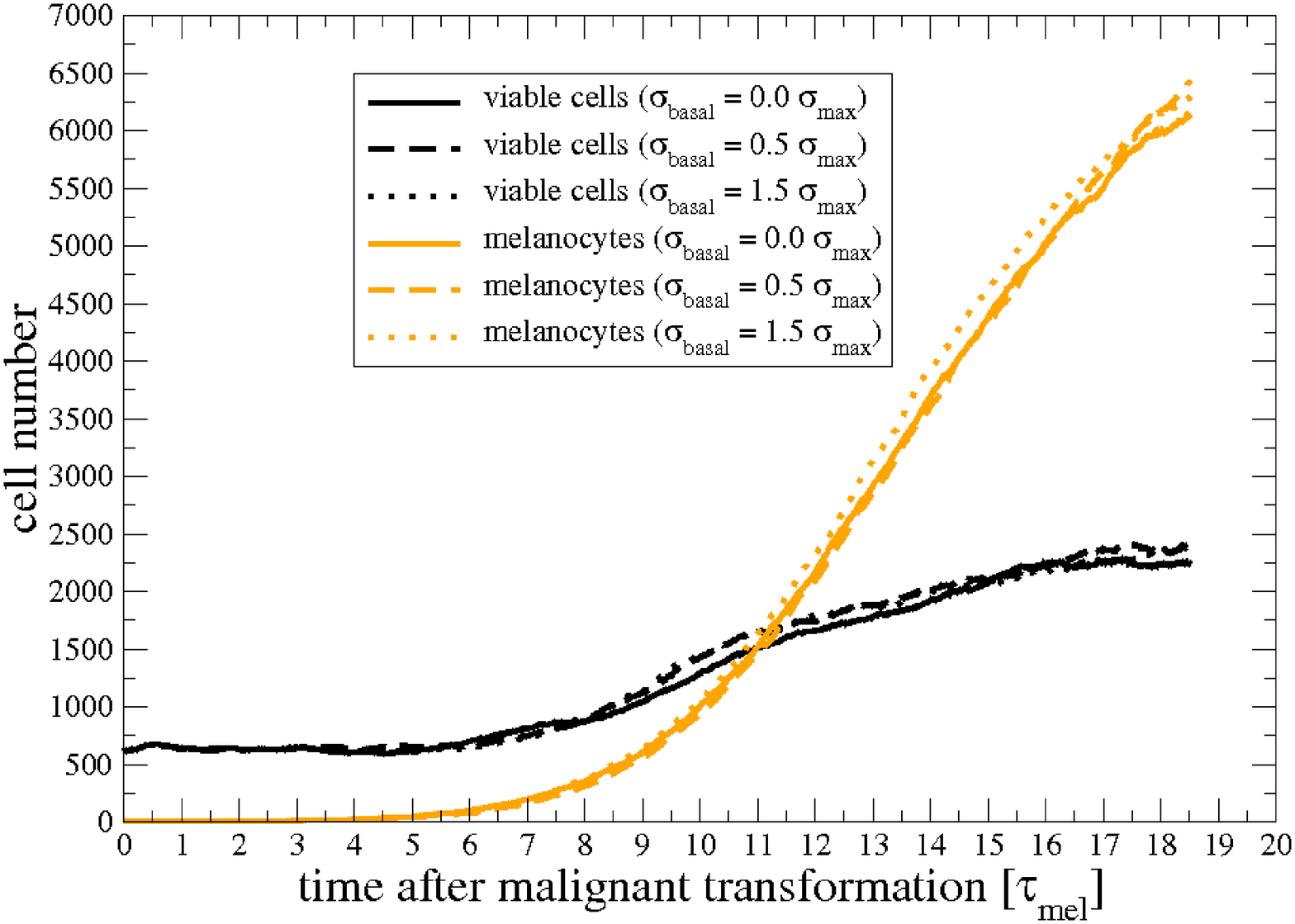}
\end{minipage}
&
\begin{minipage}{0.5\linewidth}
\includegraphics[height=6cm]{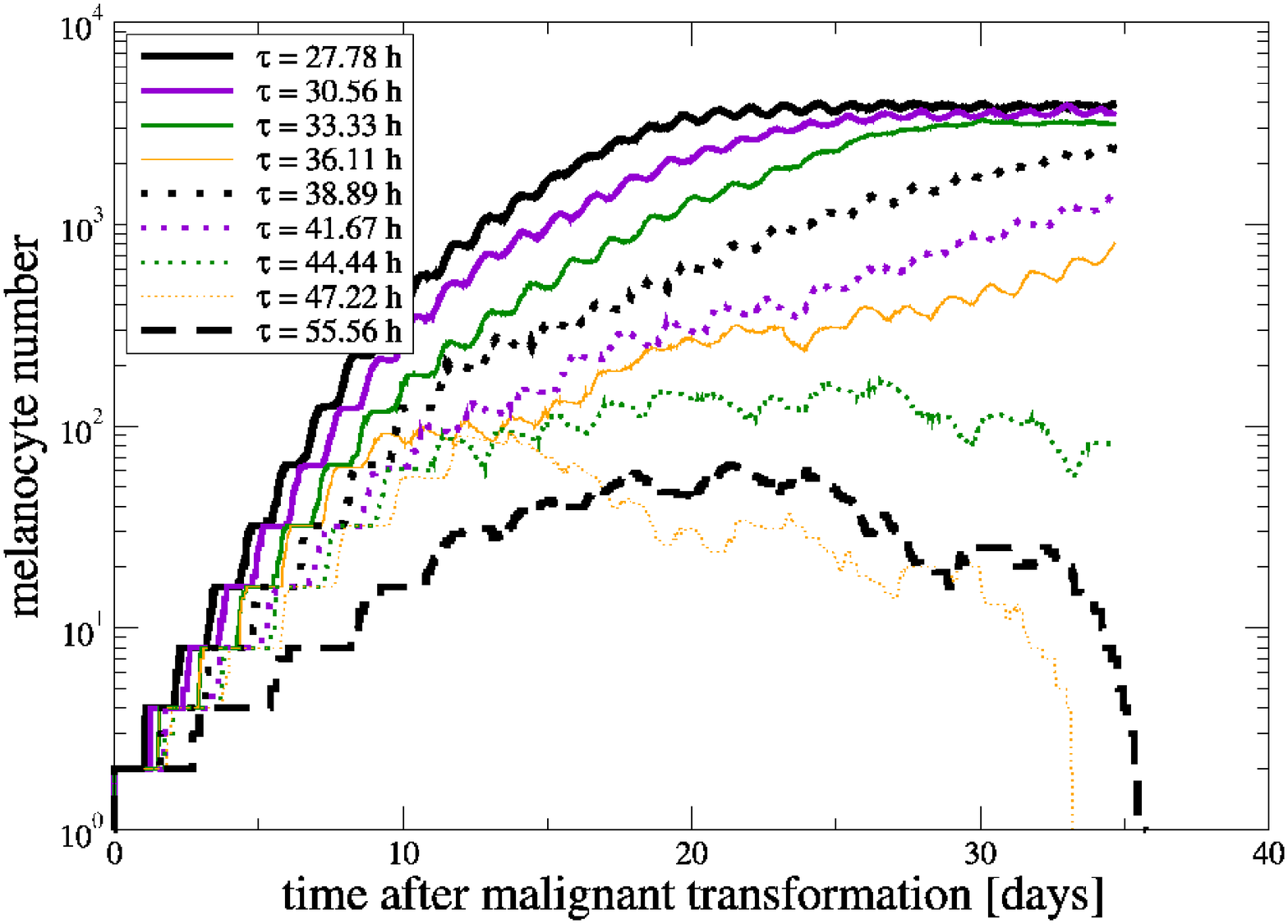}
\end{minipage}
\end{tabular}
\caption{\label{Fbasal_adhesion}
{\bf Left}: Total number of melanocytes (grey, including necrotic and viable
melanocytes) and viable cells (black, including stem cells,
keratinocytes, and melanocytes) for different degrees of basal
adhesion versus time -- expressed in units of the cancerous melanocyte
cycle time. For the assumed melanocyte cycle time of $\tau_{\rm
mel}= (15.0 \pm 2.0)$ h and other parameters chosen as in
table~\ref{Tparameters}, the basal anchorage has no significant effect
on the overall dynamics.
{\bf Right}: With slower melanocyte proliferation and
completely absent adhesion to the basal layer, a
parameter regime can be found where melanoma do not persist within the
steady-state flow equilibrium.  Although melanocyte proliferation is
still much faster than keratinocyte proliferation, the system is
sensitive to stochastic effects, as is also indicated in the
disturbed order (some curves intersect). The slowly-damped
oscillations stem from the small standard deviations of melanocyte cell cycle
durations (2 h in every run). 
}
\end{figure}

Another question is how the degree of anchorage
to the basal layer influences the ability of cancerous melanocytes to
persist within the skin. It is well-known that most human melanoma cell lines 
have decreased or no expression of cadherins and exhibit a decreased
ability to adhere to keratinocytes \cite{tang1994}. 
Therefore, this question is especially interesting from a clinical
point of view.
At first, we suspected that increased basal adhesion would lead
to an increased fraction of melanocytes bound to the basal membrane
and thereby a smaller fraction that is shed to regions where
the nutrient supply falls below necrosis-inducing levels. Thus, the
total number of melanocytes should decrease with decreasing
anchorage. 
In order to test this, a single (non-proliferating) melanocyte was
placed at the basal layer in the centre of the computational domain,
and the system was evolved until flow equilibrium 
was established. Then, the melanocyte was turned cancerous by suddenly
allowing for proliferation with a much larger rate than
keratinocytes. In addition, we concomitantly reduced the anchorage to the basal
layer. Starting from experience with multicellular tumour
spheroids \cite{schaller2005a}, we assumed the cycle time of cancerous
melanocytes to be in the order of 15 hours.
Surprisingly, it turned out that the overall growth dynamics was
hardly dependent on the anchorage to the basal layer, see
figure~\ref{Fbasal_adhesion} left panel. 
Initially, the growth of melanocytes follows an exponential growth
law, which is soon slowed down since the melanocytes 
reach distant regions from the basal layer, where nutrient support is
poor. Since due to nutrient depletion the total number of viable
cells already indicates saturation, also the total number of
melanocytes must saturate eventually. 
Even with no adhesion to the basal membrane, comparable numbers
of tumour cells were produced. Direct observation of the cross-sections
(not shown) revealed the reason: With the given melanocyte proliferation 
rate of 15 hours, exponential growth was always faster than the
epidermal flow induced by the turnover on the basal layer. 

Consequently, we varied the
proliferation rate of the cancerous cells in combination with complete
loss of basal membrane anchorage (see figure~\ref{Fbasal_adhesion}
right panel). It turned out that there is a
region of proliferation rates, where the melanocytes do not persist
within the epidermis. This region is separated from the region
of melanoma persistence by a comparably large domain where stochastic
effects become important. Interestingly, in this case the
period of coexistence of healthy skin and transformed cells may be
remarkably long, which may give time for further malignant
transformations in the realistic epidermis. It should be stressed that
in this region the melanocyte proliferation rate is still much larger
than the keratinocyte proliferation (their cycle prolongation is
active for small melanoma). In addition, in the absence of death processes the growth law
of keratinocytes follows the equation $\dot{n}_{\rm ker} = 16
\alpha_{\rm stem} n_{\rm stem}$, if one neglects the retardation
induced by the four transient keratinocyte proliferations. 
This leads to linear growth only (with a fixed number of stem cells
$n_{\rm stem}$), whereas the growth law of malignant melanocytes
will be exponential 
$n_{\rm mel} = n_{\rm mel}^0 e^{\alpha_{\rm mel} t}$ in the absense of
death processes.

\begin{figure}[htb]
\begin{tabular}{cc}
\begin{minipage}{0.6\linewidth}
\includegraphics[width=11cm]{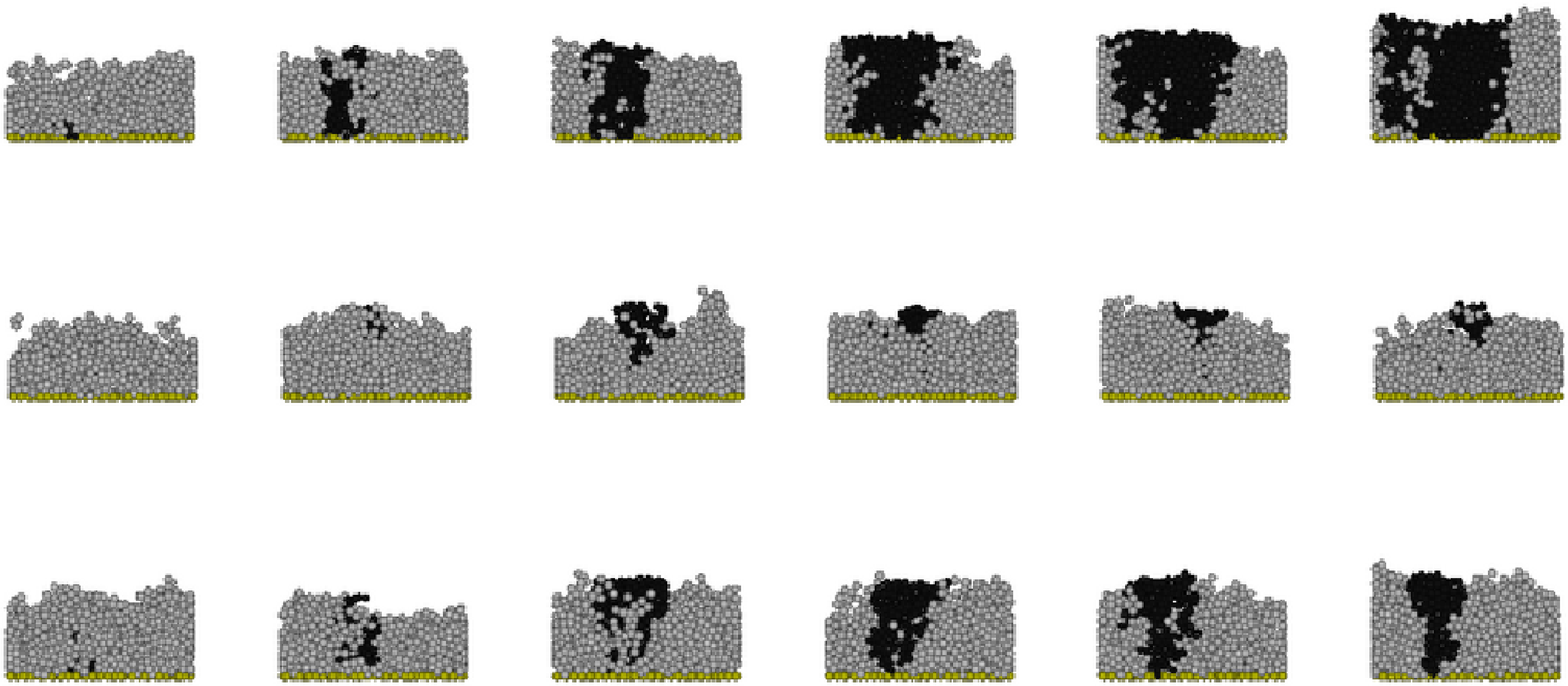}
\end{minipage}
&
\begin{minipage}{0.4\linewidth}
\includegraphics[height=6cm]{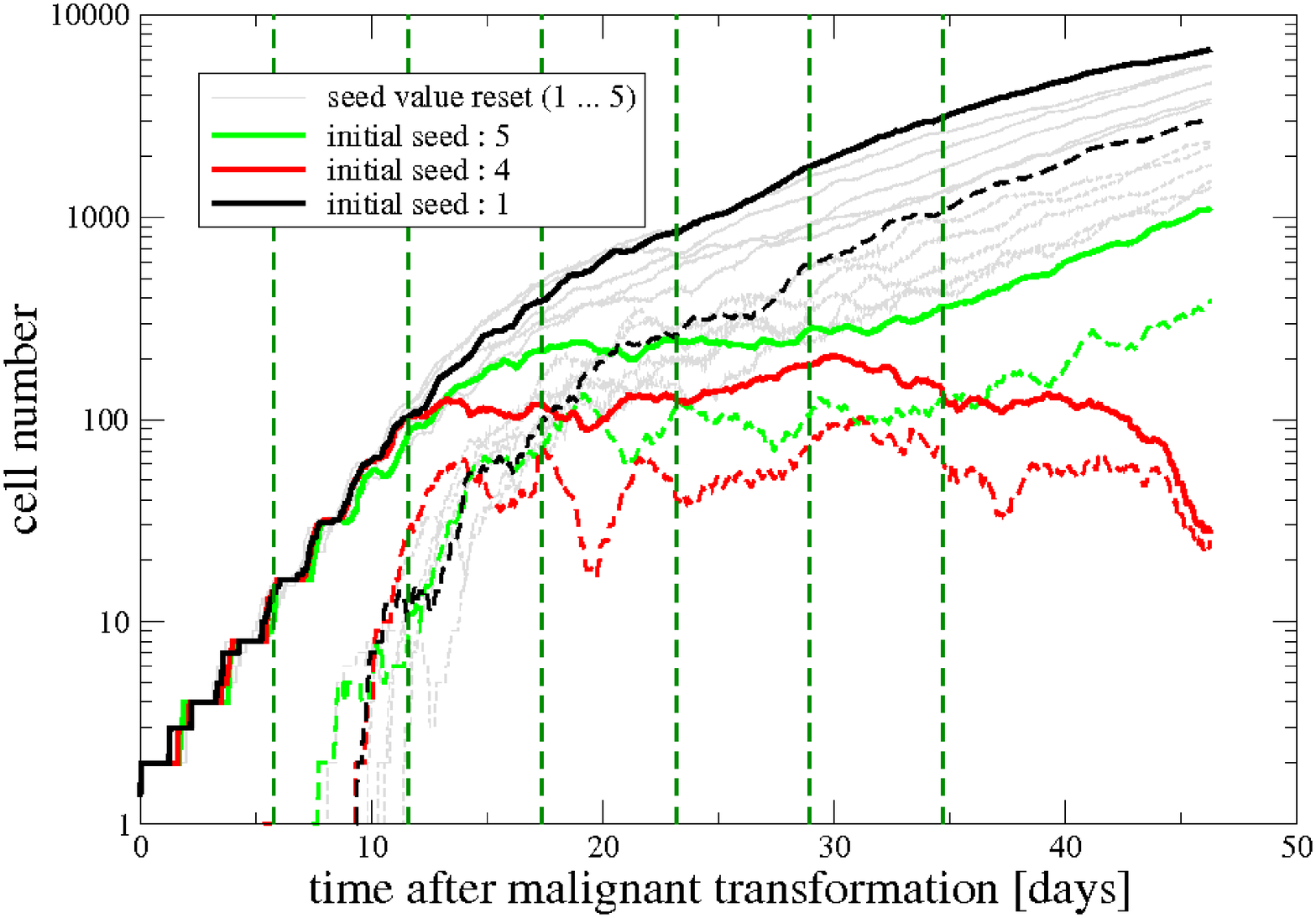}
\end{minipage}
\end{tabular}
\caption{\label{Fcoexistence}
{\bf Left}: {\em In silico} evolution of cancerous melanocytes (black)
within an epidermal population containing keratinocytes (light grey)
and stem cells (the bottom layer). The cross-sections do not
distinguish between viable and necrotic or viable and cornified cells,
respectively. The times at
which the cross-sections have been produced are marked in the right
panel with dashed lines. The first row corresponds to the seed
value 1 (thick black curves in the right panel), the second row to seed value
4 (thick curves in grey), and the last row to seed value 5 (thick
curves in light grey).
Note that the diameter of a single frame is $200 \rm\;\mu m$ only,
such that the figure represents the state before clinical manifestation.
{\bf Right}: Total number of melanocytes (thick lines) and necrotic
melanocyte subpopulation (associated dashed curves emerging around day 10)
after the malignant transformation. For the thick lines, the different
shades of grey correspond to different initial seed values (with
otherwise equal parameters) of the random number
generator. Consequently, these simulations also already include a changed
environment for the melanocyte at the time of its malignant
transformation. Note that for seed value 4 there are hardly any viable
melanocytes left after 45 days (dashed and solid curves combine). This
is different if the stochastic effects do not refer to the initial
environment configuration of the cancerous melanocyte (thin
curves in light grey). 
}
\end{figure}

We further examined the region that separates melanoma persistence and
complete shed-off of cancerous melanocytes by changing the 
melanocyte cycle time to $\tau_{\rm mel}= (44.44 \pm 5.56)$ h. 
One finds that the usual spherical form one observes for 
{\em in vitro} tumour spheroids is considerably deformed for this
system to cylinder-shaped or cone-shaped, compare
figure~\ref{Fcoexistence} left panel. This is due to the pre-existent
flow-equilibrium of the surrounding tissue and the effective
one-dimensional diffusional constraint.
Note also that the boundaries of the tumours are rather
diffuse. Initially, a thin column of cancerous melanocytes is
formed. Then, in the example in figure~\ref{Fcoexistence}, left panel,
first row, the melanocytes can persist within the
life-sustaining zone until their growth velocity outweighs the
upward-directed flow velocity and direct contact with the basal
membrane is re-established. Afterwards, in the middle of the column of
cancerous cells the upward forces are decreased,
since for the interior cells there is no direct contact with
keratinocytes moving upwards.
In the simulations in figure~\ref{Fcoexistence} left panel, the
thickness of the epidermis increases in those simulations where the
tumour has re-established contact with the basal membrane. This is due
to the displacement of keratinocytes -- which are constrained in $x$
and $y$ dimensions -- and also to the loss of the protective cornified
layer, which leads to enlarged keratinocyte proliferation rates. 
It may be speculated that the cross-sections correspond to initial
stages of a highly aggressive nodular melanoma \cite{moncrieff2001}
that has not yet become clinically manifest.
It may also be hypothesised that the micrometastases sometimes observed around
primary melanoma in skin may correspond to branches of melanoma clones
that have separated from the main clone during the upward
flow. Interestingly, the shapes of these structures appear to be dynamically
changing in these initial phases.

Using different initial seed values for the random number generator, we have
performed several simulations with otherwise equal parameters. It
turns out that completely different outcomes may occur in this region
of melanocyte proliferation rates (figure~\ref{Fcoexistence} left
panel and thick curves in the right panel). 
The stochastic effects result from stochastic forces, the randomly
chosen mitotic direction, and the randomly distributed duration times
of the cell cycle. 
In this {\em in silico} experiment, the different seed value did
already lead to different configurations before the malignant
transformation. More specific, the initial conditions for the growth
of cancerous melanocytes had also been varied by employing stochastic
elements before.
In order to separate these effects, we started another
series of simulations with equal initial seed values. In contrast to
the previous simulations, the seed value of the random number
generators was reset to different values right at the 
time of the malignant transformation. Thus, the initial environment of
the cancerous melanocyte was the same in these
simulations. It turned out that the variance of the
outcomes narrowed considerably (thin grey curves in
figure~\ref{Fcoexistence} right panel) but still exhibit large
variations in the cell number (logarithmic plot). 
Thus, it can be concluded that the initial environment of cancerous
melanocytes contributes significantly to the final outcome. Note that
this does not only refer to the spatial cellular position, but
also to the local proliferative state and thereby to the local upward
flow velocity: The upward drag forces will be larger if the cancerous
cell is surrounded by many proliferating keratinocytes with a net upward flow
velocity. 

In conclusion, stochastic effects generally play an important role in the
initial phases of {\em in silico} melanoma development, since for the
small cell numbers in the initial phases, they do not
average out completely. In addition, their secondary consequences,
i.~e., the variation of the initial local environment by stochastic
influences, are relevant. 

\section{Model parameters}\label{Smp}

Reasonable dynamics has been achieved with the parameters in
table~\ref{Tparameters}.  
\begin{table}
\begin{tabular}{c|c|c}
parameter & value & comment\\
\hline\hline
ECM viscosity $\eta$ & 0.001 kg/($\mu$m s) & \cite{schaller2005a,galle2005a}\\
adhesion energy density $\sigma^{\rm max}$ & 0.0001 $\rm Nm/m^2$ & \cite{chu2004}\\
minimum anchorage $\Sigma^{\rm min}$ & 0.00001 pJ & estimate\\
receptor loss rate $\alpha$ & 0.00001/s & estimate\\
tangential friction coefficient $\gamma_\parallel$ & 0.1 $\cdot 10^{12} \rm Ns/m^3$ & \cite{galle2005a}\\
stochastic force coefficient $\xi$ & 0.001 $\cdot 10^{-6} \rm kg\;
m/s^{3/2}$ & D = 0.0001 $\rm\mu m^2/s$ \cite{schaller2005a,galle2005a}\\
M-phase duration $\tau^{\rm (m)}$ & (1.0 $\pm$ 0.25) h & \cite{schaller2005a}\\
$\rm S/G_2$-phase duration $\tau^{\rm (S/G_2)}$ & (5.0 $\pm$ 2.0) h & \cite{schaller2005a}\\
keratinocyte $\rm G_0$-phase prolongation $\tau^{\rm (G_0)}$ & (138.9 $\pm$ 138.9) h & \cite{barthel2000}\\
shortest observed keratinocyte cycletime $\tau^{\rm min}$ & 15 h & \cite{schaller2005a,barthel2000}\\
pre-mitotic cell radius $R^{\rm (m)}$ & 5.0 $\mu$m & \cite{norlen2004}\\
cell elastic modulus $E_i$ & 750 Pa &\cite{mahaffy2000}\\
cell Poisson ratio $\nu_i$ & 1/3 &\cite{maniotis1997} \\
melanocyte glucose uptake rate $\lambda^{\rm mel}$ & 150.0 amol/(cell
s) & \cite{wehrle2000}\\
keratinocyte glucose uptake rate $\lambda^{\rm ker}$ & 10.0 amol/(cell
s) & estimate\\
critical water concentration $U_{\rm H_20}^{\rm crit}$ & 90.0 \% &
\cite{kasting2003}\\
maximum water concentration $U_{\rm H_20}^{\rm bound}$ & 100.0 \% &
by definition\\
critical glucose concentration $U_{\rm gluc}^{\rm crit}$ & 1.0 mM & \cite{freyer1986a}\\
water diffusivity $D_{\rm H_20}^{\rm strat. med.}$ & $1000.0\;\rm \mu m^2/s$& \cite{chenevert1991,livingston1995}\\
water diffusivity $D_{\rm H_20}^{\rm strat. corn.}$ & $0.2 \;\rm\mu m^2/s$ & \cite{schwindt1998,bashir2001}\\
glucose diffusivity $D_{\rm gluc}^{\rm tissue}$ & $256.0 \;\rm \mu
m^2/s$ & \cite{tuchin2001}\\
glucose boundary concentration $U_{\rm gluc}^{\rm bound}$ & 5.0 mM & \cite{carvalho2004}\\
stem cell basal adhesion energy density $\sigma^{\rm basal}$ & $2\sigma^{\rm max}$ & estimate\\
\end{tabular}
\caption{\label{Tparameters}
Model parameters
have been estimated from independent experiments or they have been
varied as fit parameters. Parameters not included in the table have
been also varied and are discussed in separate sections of this article.} 
\end{table}

The viscosity of the extracellular matrix $\eta$ determines the friction on
loosely bound cells. Large viscosities lead to increased
friction. Since viscous friction due to the cytoskeleton is
assumed to be small ($\gamma_\perp \approx 0$, compare appendix \ref{ASeom}) -- also dominates
friction in directions normal to the cell contact surfaces. Thereby,
also the initial speed of cell division in M-phase 
is dominantly dependent on $\eta$. 
As long as the force relaxation occurs on a shorter
timescale than the total cell doubling time, this does not have macroscopic
effects on the evolution of the tissue.
This is different when $\gamma_\perp$ does not vanish. If it has the
same order of magnitude as $\gamma_\parallel$, it will dominate the
contribution inflicted by the viscosity $\eta$. However, if the
magnitude of the total drag force coefficient does not change (marked
by $\gamma^2 = \gamma_\perp^2 + \gamma_\parallel^2$), we have found
by comparing the three extreme cases (that is,
$\gamma_\perp=0, \gamma_\parallel=\gamma$ and 
$\gamma_\perp=\gamma, \gamma_\parallel=0$ and
$\gamma_\perp=\gamma_\parallel=\gamma/\sqrt{2}$) that the differences
in the overall population dynamics are rather small. It may be
speculated that this is because in the present calculations the
relaxation speed has no direct back-reaction on the number of cells,
as in contrast to \cite{galle2005a,schaller2005a} contact inhibition
has not been included in the model.
As here absence of perpendicular friction has been assumed, the
tangential friction coefficient $\gamma_\parallel$ dominantly
determines the speed of relaxation within the tissue. The chosen value
led to reasonable dynamics and has been estimated from \cite{galle2005a}.

The adhesion energy density $\sigma^{\rm max}$ determines the cell-cell
equilibrium distance and the binding strength, which was a marker for
the removal of necrotic or cornified cells. Generally, 
this value will in reality be time-dependent, compare also the
discussion at the end of subsection~\ref{SScmv}.
Therefore, the binding energy density has been derived from the
observed equilibrium distance \cite{chu2004} solving (\ref{Ejkr_model})
instead. With this procedure, the equilibrium distances are in a
physiological regime. Note that larger adhesion will lead to smaller
equilibrium distances (with moderately increased contact surfaces and
drag forces) but also to longer persistence times of dead cells, which
results in an increased thickness of the {\em stratum corneum}. However, due
to equation (\ref{Erecligrate}) this latter effect only enters
logarithmically. When both the adhesion energy $\sigma^{\rm
max}$ and the minimum anchorage $\Sigma^{\rm min}$ are decreased, one
will still have to decrease the maximum stepsize in the numerical
solution to maintain the level of accuracy. This is due to the fact
that for decreased adhesion, the equilibrium distance and the contact
distance $d_{ij}^{\rm contact}=R_i+R_j$ are closer together.

The equilibrium thickness of the cornified layer is strongly dependent
on the receptor loss rate $\alpha$ and the minimum anchorage
$\Sigma^{\rm min}$. In addition, it will be 
sensitive to the cycle times of stem cells and keratinocytes, since these
determine the number of keratinocytes finally undergoing
cornification.

The elastic parameters correspond to approximate physiologic values for cells
\cite{mahaffy2000,maniotis1997,galle2005a}. However, it is known that --
depending on the cell type and individual cytoskeleton -- significant
deviations may occur. With the given drag forces, mechanical relaxation
occurs on a shorter scale than the cell cycle times, such that changes
in physiologic windows have only small macroscopic consequences. It
should be noted however that already for moderately changed Young moduli (and/or
reduced Poisson moduli) the equilibrium distance between cells will be
shifted, which might make decreased maximum spatial stepsizes
necessary in the numerical solution to avoid unphysiological losses of
contact.

As has already been discussed above, the stochastic elements may have
significant influence on melanoma development. These can be divided
into stochastic forces, randomly chosen durations of the cell cycle
phases, and the random direction of mitosis.

Stochastic forces contribute to the detachment of cornified and
necrotic cells, since these do neither advance in the cell cycle nor
proliferate. We have found that small variations in the strength of
stochastic forces in physiologic 
regimes only change the fluctuations in the epidermal thickness around
the unchanged equilibrium value.
On a technical level, the existence of a planar basal layer in
combination with 
completely absent stochastic forces sometimes led to planar cell
configurations at the basal layer, which is unfavourable for the
Delaunay triangulation \cite{schaller2004}. As may be expected,
considerably larger stochastic forces have a strong influence on the thickness
of the {\em stratum corneum}, since loosely bound cells are removed
much faster and the protective layer is lost easily. This in turn leads
to loss of water and on-going reactions of keratinocytes that leave
$G_0$-phase.

The values of the durations of M-phase $\tau^{\rm (m)}$, the
$S/G_2$-phase $\tau^{\rm S/G_2}$ and the prolongation of the cell
cycle $\tau^{\rm (G_0)}$ influence the relative distribution
of cells within the cell cycle, whereas the sum of their squared
widths primarily 
determines the speed of desynchronisation of cell division (compare
figures~\ref{Fbasal_adhesion} and~\ref{Fcoexistence} right panels). 
Due to missing data, the durations of these cellular
states have been fixed from a previous publication
\cite{schaller2005a}. The shortest observed cycle 
time determines the proliferation time for keratinocytes when the water
concentration is below the critical threshold and has been estimated
from experimental observations \cite{potten2000}. The system is most
sensitive to the $\rm G_0$-phase prolongation time $\tau^{\rm G_0}$,
which has been estimated from \cite{barthel2000} to yield
reasonable dynamics.

Without the modelling constraint that on division of keratinocyte
stem cells, only the upper cell becomes a differentiating
keratinocyte, the basal layer would loose more and more stem cells in
the model. In other cell divisions, the simple assumption of a randomly
distributed initial mitotic direction did not lead to numerical
artifacts. However, it can be expected that the configuration of the
neighbour cells soon changes the initial direction of the mitotic
doublet. 

The average cell volume of keratinocytes varies from $425 \rm\; \mu m^3$
for cornified cells to $800 \rm\; \mu m^3$ for {\em stratum granulosum}
keratinocytes \cite{norlen2004}. Therefore, with the intrinsic
assumption of spherical shape, the maximum cell radius has been fixed
to $R^{\rm (m)} = 5.0 \rm \;\mu m$, which also influences the
time-dependent target volume. Note however, that within the 
{\em stratum corneum} the cornified cells flatten considerably and the
realistic intrinsic cell shape cannot be regarded as spherical anymore.

The glucose uptake rate for cancerous melanocytes $\lambda^{\rm
(mel)}$ has been chosen considerably larger than the glucose uptake
rate of keratinocytes $\lambda^{\rm (ker)}$. This is motivated by the
observation that cancerous cells have a considerably increased metabolism. The actual
values are in the range observed for other tumour cells \cite{wehrle2000}. 
The minimum nutrient concentration $U_{\rm gluc}^{\rm crit}$, below
which for melanocytes necrosis is induced, has been chosen to be in
the order of $1 \rm\; mM$, since necrosis of cancer cells becomes visible at
these nutrient concentrations {\em in vitro}
\cite{freyer1986a,freyer1986b}. Thereby, the combination of melanocyte
nutrient uptake rate and minimum glucose concentration define a
region, where melanocytes can survive.

For simplicity we have assumed that as a net effect the cells do not
consume or secrete mobile water. A possible model extension could
incorporate such effects by including cellular swelling during
hydration. The critical mobile water concentration $U^{\rm crit}_{\rm
H_20}$ has been adjusted to obtain a reasonable equilibrium thickness
of the {\em stratum medium} with $\order{5}$ cell layers.  

The apparent diffusivity of the mobile water $D_{\rm H_20}^{\rm strat. med.}$ in
{\em stratum medium} as well as in {\em stratum corneum}
$D_{\rm H_20}^{\rm strat. corn.}$ has been determined experimentally
by various studies. Though strong variances exist, all of them
predict a strong decline of the apparent diffusion
coefficient \cite{kasting2003,bashir2001,schwindt1998}. 
Roughly speaking, the local
water diffusion coefficients influence the gradient of the mobile
water concentration: Large diffusion coefficients correspond to a small
gradient. Therefore, for an intact {\em stratum corneum} the water
concentration is approximately constant throughout the 
{\em stratum medium} and then falls rapidly, compare also figure
~\ref{Ftape_stripping} right panel.

The same general features hold true for the glucose diffusion coefficient
$D_{\rm gluc}^{\rm tissue}$, which has specifically been determined
for the human skin \cite{tuchin2001}.
The glucose concentration at the basal layer 
$U_{\rm gluc}^{\rm bound}$ has been fixed to values 
that are normal for blood \cite{carvalho2004}. 
However, it should be noted that in reality the blood
glucose concentration may vary significantly -- for example after a
meal. Since within the model for normal parameter sets anoikis is the
predominant pathway for keratinocytes and dominantly the cancerous
melanocytes consume glucose at large rates in the model, the glucose
concentration strongly influences the chances of melanocyte survival
here. An improved model could for example include an intracellular
glucose reservoir to average out the time-dependent supply.

In order not to loose stem cells at the basal layer migrating upwards
to the {\em stratum corneum}, the basal adhesion energy has been
chosen to be twice the maximum adhesion energy density $\sigma^{\rm
max}$. This did suffice to disable loss of stem cells. For
non-proliferating melanocytes, the basal adhesion has been
chosen similarly. 


\section{Discussion}\label{Sd}

It had been demonstrated already in \cite{schaller2005a} that with the aid of
kinetic and dynamic weighted Delaunay triangulations agent-based
models can treat up to $10^5 \ldots 10^6$ cells. In the present
contribution, it has been shown that with a more complete treatment of
the equations of motion, such models can still handle $10^4 \ldots
10^5$ cells. 

Apart from these technicalities, from a biological point of view a
diffusible substance can serve as a moderator on cellular
proliferation in the epithelium. The parameters used do not contradict
that a simple candidate of this substance could be the mobile water in
the tissue. 
The homoeostasis was found to be roughly stable against perturbations such as
tape-stripping experiments, which can serve as a sanity check on the
model implementation.

Independently, the consequences of a varying basal adhesion of
cancerous melanocytes have been studied. It turned out that these
are strongly interlinked with the balance of proliferative
melanocyte and keratinocyte activities.
In addition, it has been shown that in some regions of parameter
space, stochastic effects and especially their consequences on the
initial state on the environment play an important role
in the {\em in silico} representation of melanoma growth. 
Evidently, the model behaviour has been found under the precondition
of several explicit and implicit approximations. 
These do of course limit the generality of the model and we want to
summarise some shortcomings of the model below:

From our point of view, a significant macroscopic shortcoming
of our approach is the failure of the model to explain the reduced
thickness of the  {\em stratum corneum}. This is at least partly due
to the fact that the inherent cell shape is spherical, whereas
cornified cells flatten and form polarised adhesive bindings
\cite{montagna1992}. In reality, this will lead to a greater stability
of the {\em stratum corneum} in comparison to the model, which would
also imply a smoother evolution around the steady-state flow
equilibrium than exhibited in figure~\ref{Ftape_stripping} left panel.
Possibly, choosing ellipsoids in
contrast to spheres as the intrinsic cell shape \cite{dallon2004}
may provide an alternative. Another possibility would be to use
boundary-based models such as e.~g.~the extended Potts model
\cite{savill2003}.

From the theoretical point of view, the model could be significantly
improved by deriving a contact model valid for two-body interactions
that also include non-normal forces and do not underlie the
constraints of only small cell deformations. Also, for {\em in vitro}
cell populations that are not fixed to a substrate, the effects of
torque may become important.
These refined theories however require much better experimental
resolution than currently provided. It appears questionable whether
centre-based models are able to cope with the increasing degree of
complexity resulting from these improvements.

The basal layer has been approximated with a plane boundary condition
in this article. Its replacement by a corrugated structure would 
significantly enlarge the region
where water and nutrients are provided in abundance and thereby lead
to a far greater cell reservoir that is able to start a proliferative
response in case of injury. It may be speculated that this is one of
the reasons why the ruffled basal layer has developed in skin.
In addition, one would expect that a ruffled basal layer will also
lead to a ruffled skin surface.
Especially for the clinical question of melanoma invasion depth, the
plane boundary condition should be replaced by a boundary that can be
penetrated by malignant melanocytes. This would allow to study the
time-course of initial invasion and to compare the invasion depth with
clinical melanoma classifications.

The dynamics of the nutrients and of water has been described with a
reaction-diffusion approach here. However, due to the cellular
movement, there will also be a convective and a transport contribution
that is completely neglected in the current simulations. With the
large diffusion coefficient for water and nutrients in viable tissues,
this approximation is presumably valid within the viable layers but
may be questionable in the {\em stratum corneum}.
Note that the polarised structure of 
the cornified cells in the {\em stratum corneum} may also give rise to
non-isotropic diffusion, where the diffusion coefficient is not a
scalar value anymore. To a first approximation however, this effect may
be well absorbed into the apparent diffusion coefficient as is done in
the experimental measurements.

The cell cycle has been approximated here by a small number of
internal cellular states only. It may also be questioned whether a
subdivision into discrete substates makes sense. One may also
expect a much smoother reaction of the epidermis to the removal of all
keratinocytes if transition into and out of $\rm G_0$-phase would not
depend on a threshold water concentration, but would be determined by
transition probabilities that may continuously depend on the water
concentration. This may be judged with quantified experimental data.

The model also uses comparably many parameters but all of them have a
distinct physical counterpart. This makes it in principle possible to
determine these parameters by independent experiments.
Despite of all the previously-mentioned shortcomings (most of these
being valid for lattice-based approaches as well), off-lattice
agent-based models also have important advantages over most
lattice models: They have the intrinsic
potential to use physical (realistic) parameters with a moderate
increase in computational effort. This opens the
possibility to gain knowledge about the system by falsifying the model
using independent experiments. 
Therefore, quantified experiments on well-defined experimental systems
are of urgent interest to constrain the uncontrolled growth in the
number of theoretical models on cellular tissue. 


\section{Acknowledgements}\label{Sa}

G.~S.~is indebted to J.~Galle and T.~Beyer for valuable discussions on
contact models, physiologic parameters, and numerical algorithms. 
FIAS is supported by the ALTANA AG.


\appendix


\section{The JKR contact model}\label{ASjkr}

Already the dynamics of rigid bodies in contact is a difficult
problem, as the local geometry at the contact region will strongly
influence the involved forces. Therefore, most contact models applied in
practice are not motivated by microscopic assumptions but rather mimic
the realistic behaviour.

The JKR-model includes elastic and adhesive (but not viscous)
interaction of solid spheres. It is often used in a
biological context to estimate cellular parameters from experimental
observations (JKR-test, \cite{verdier2003}). Thus, one can at least on
short time scales hope, that even though the
parameters derived from such measurements \cite{moy1999} will not
yield a correct  description of the cytoskeleton (which is known to be
viscoplastic), their usage in the model will at least lead to dynamics
similar to that observed in the experiments.

\begin{figure}[htb]
\begin{tabular}{cc}
\begin{minipage}{0.4\linewidth}
\includegraphics[height=5cm]{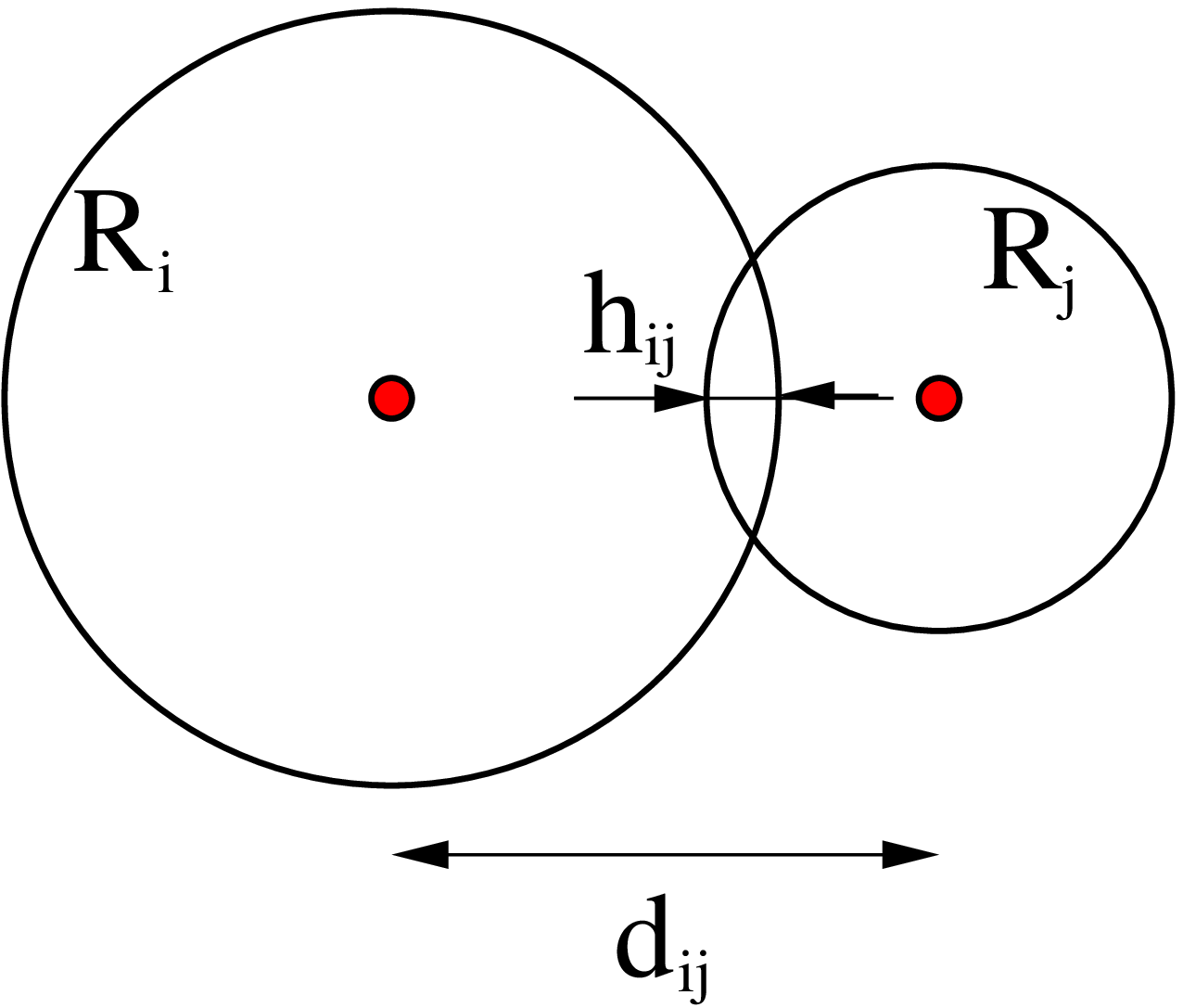}
\end{minipage}
&
\begin{minipage}{0.6\linewidth}
\includegraphics[height=6cm]{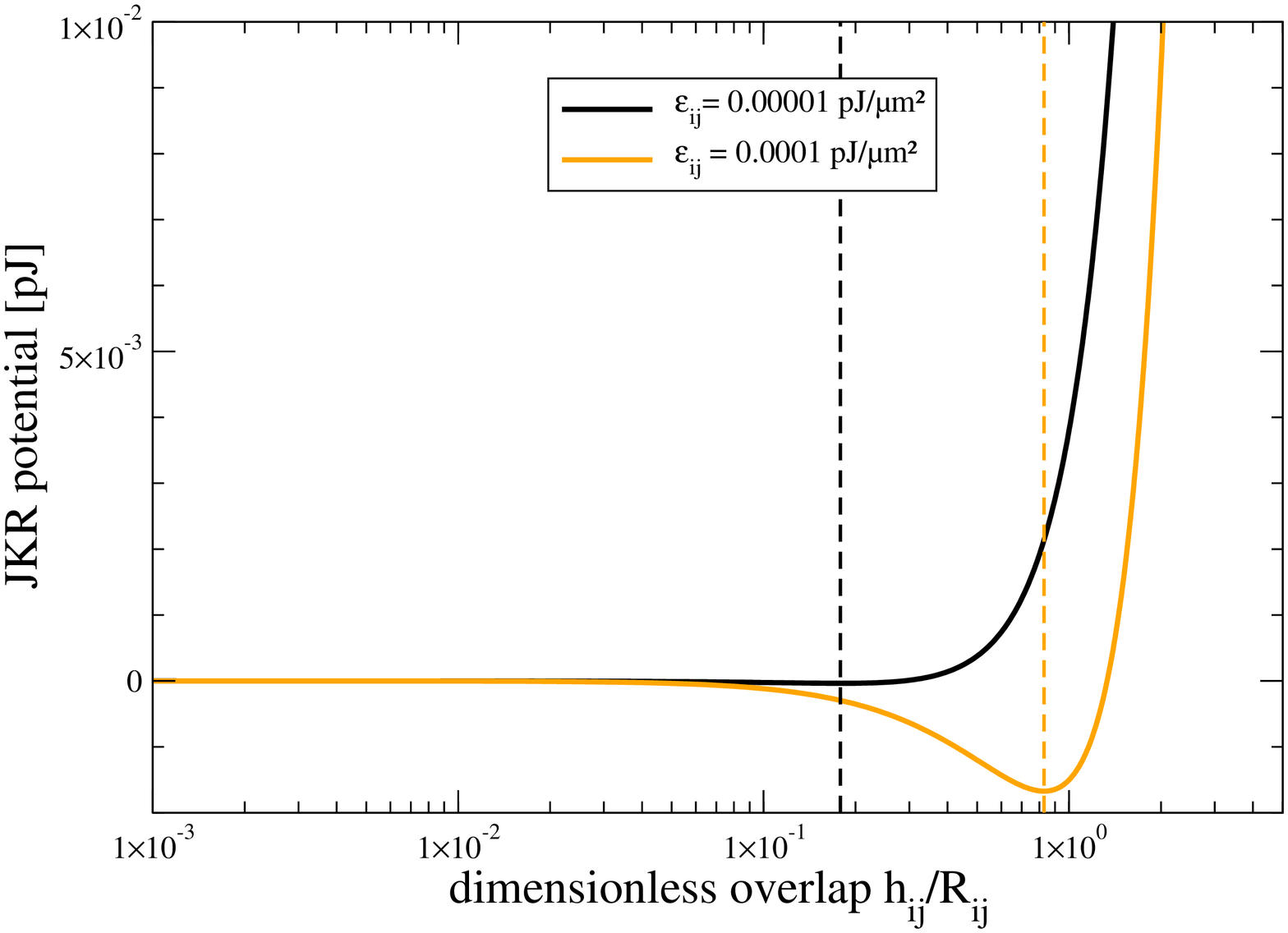}
\end{minipage}
\end{tabular}
\caption{\label{Fjkr_model}
Illustration of the interaction model. The dynamics of two elastic
and adhesive spheres in contact can be described with the
JKR-potential. 
{\bf Left}: The virtual sphere overlap $h_{ij}$ 
can be calculated from the sphere distance 
$d_{ij} = \abs{\Vektor{x}_i - \Vektor{x}_j}$ and
equilibrium radii $R_i$ and $R_j$. The circle constructed by the
intersection of the sphere surfaces does not directly define the
contact radius $a_{ij}$, since in real world scenarios, the spheres would
evidently deform (not shown).
{\bf Right}:
The existence
of adhesive forces gives rise to bound states (right, minima at dashed
lines). Their position and depth strongly depends on the parameters
$\sigma_{ij}$ and $K_{ij}$. Note that the potential does not diverge
at $h_{ij}/R_{ij}=2+2R_{\rm min}/R_{\rm max}$ (complete overlap). The curves on
the right have been computed using the following (physiological) values 
$K_{ij} = 1000$ Pa, $R_{ij} = 2.5 \rm\;\mu m$.
}
\end{figure}

The characteristics of the JKR contact model relevant for our
considerations can be summarised as follows: 
Two spheres $i$ and $j$ placed at positions $\Vektor{x}_i$ and 
$\Vektor{x}_j$, having radii $R_{i/j}$, Young moduli $E_{i/j}$,
Poisson moduli $\nu_{i/j}$, and contact surface energy density
$\sigma_{ij}$ underlie the interaction force \cite{johnson1971}
\bea\label{Ejkr_fullmodel}
F_{ij}^{\rm JKR} &=& 
\left[\frac{K_{ij} a_{ij}^3}{R_{ij}}
- \sqrt{6\pi\sigma_{ij} K_{ij} a_{ij}^3}
\right]\,,
\eea
where $a_{ij}$ denotes the radius of the circular contact area between
the deformed spheres, $R_{ij}$ the reduced
radius, and $K_{ij}$ incorporates the combined elastic properties
\bea\label{Ejkr_details}
R_{ij} = \frac{R_i R_j}{R_i + R_j}\,,\qquad
K_{ij} = \frac{1}{\frac{3}{4}
\left(\frac{1-\nu_i^2}{E_i} + \frac{1-\nu_j^2}{E_j}\right)}\,.
\eea
For vanishing adhesive properties ($\sigma_{ij}=0$) one
recovers the purely elastic Hertz model \cite{hertz1882,landau1959}.
The contact radius $a_{ij}$ is related to the indentation or overlap 
(see figure~\ref{Fjkr_model} left panel) 
\mbox{$h_{ij} = R_i + R_j -\abs{\Vektor{x}_i - \Vektor{x}_j}$}
via \cite{johnson1971,brilliantov2004}
\bea\label{Eambiguity}
h_{ij} = \frac{a_{ij}^2}{R_{ij}} - \sqrt{\frac{8\pi\sigma_{ij}}{3
K_{ij}}} \sqrt{a_{ij}}\,,
\eea
which may have -- depending on the value of $h_{ij}$ -- none, one, or
two solutions with $a_{ij}>0$. For relatively small adhesion
$\sigma_{ij}/(K_{ij} R_{ij}) \ll 1$, the second term on the right hand
side can be neglected, and the solution $a_{ij}\approx\sqrt{h_{ij} R_{ij}}$
can be inserted into equation (\ref{Ejkr_fullmodel}) to yield an approximate
force-distance relationship \cite{brilliantov2004}
\bea\label{Ejkr_model}
F_{ij}^{\rm JKR} &\approx& 
\left[K_{ij} R_{ij}^2 \left(\frac{h_{ij}}{R_{ij}}\right)^{3/2}
- \sqrt{6\pi\sigma_{ij} K_{ij} R_{ij}^3} \left(\frac{h_{ij}}{R_{ij}}\right)^{3/4}
\right]\,,
\eea
which has been used as the JKR force throughout this
article. 
The force is negative (adhesive) for small overlaps $h_{ij}$
and becomes positive (repulsive) for larger overlaps. 
Note that, independent on the approximation of small adhesion in
(\ref{Eambiguity}), the adhesive force has the maximum magnitude 
\bea\label{Emaxadh}
F_{ij}^{\rm adh} = -\frac{3}{2} \pi \sigma_{ij} R_{ij}\,,
\eea
which is also independent on the elastic cell properties and thus
allows an estimate of $\sigma_{ij}$ from cell-doublet-rupture
experiments such as e.~g.~\cite{benoit2000,chu2004}. Since in reality the spheres
underlie deformation, the resulting approximate sphere contact surface
in JKR theory  
\bea\label{Econtactsurface}
A_{ij}^{\rm JKR} = \pi a_{ij}^2 \approx \pi h_{ij} R_{ij}
\eea
is in general different from the virtual contact surface that would follow
intuitively from the sphere overlap region (figure~\ref{Fjkr_model}
left panel). The above contact surface has been chosen in the model to
make it intrinsically consistent. In the following, the short hand
notations 
$R_{\rm min}=\min\{R_i,R_j\}$ and 
$R_{\rm max}=\max\{R_i,R_j\}$ will be used with suppressed indices. 

For the approximate theory (\ref{Ejkr_model}) one can introduce a
two-body interaction potential via 
\bea
F_{ij}^{\rm JKR} = -\pdiff{V^{\rm JKR}}{d_{ij}} 
= + \pdiff{V^{\rm JKR}}{h_{ij}} 
= \frac{1}{R_{ij}} \pdiff{V^{\rm JKR}}{h_{ij}/R_{ij}}\,,
\eea
which leads for our case to
\bea
V_{ij}^{\rm JKR}\left(h_{ij}/R_{ij}\right) 
= \frac{2}{5} K_{ij} R_{ij}^3 \left(\frac{h_{ij}}{R_{ij}}\right)^{5/2}
- \frac{4}{7} \sqrt{6\pi \sigma_{ij} K_{ij} R_{ij}^5} 
\left(\frac{h_{ij}}{R_{ij}}\right)^{7/4}\,,
\eea
which is a special case of the Lennard-Jones potential (compare
figure~\ref{Fjkr_model}). However, here the parameters have either  
been linked to cellular properties that are accessible by independent
experiments or been fixed by microscopic assumptions.

\begin{figure}[htb]
\begin{tabular}{cc}
\begin{minipage}{0.4\linewidth}
\includegraphics[height=5cm]{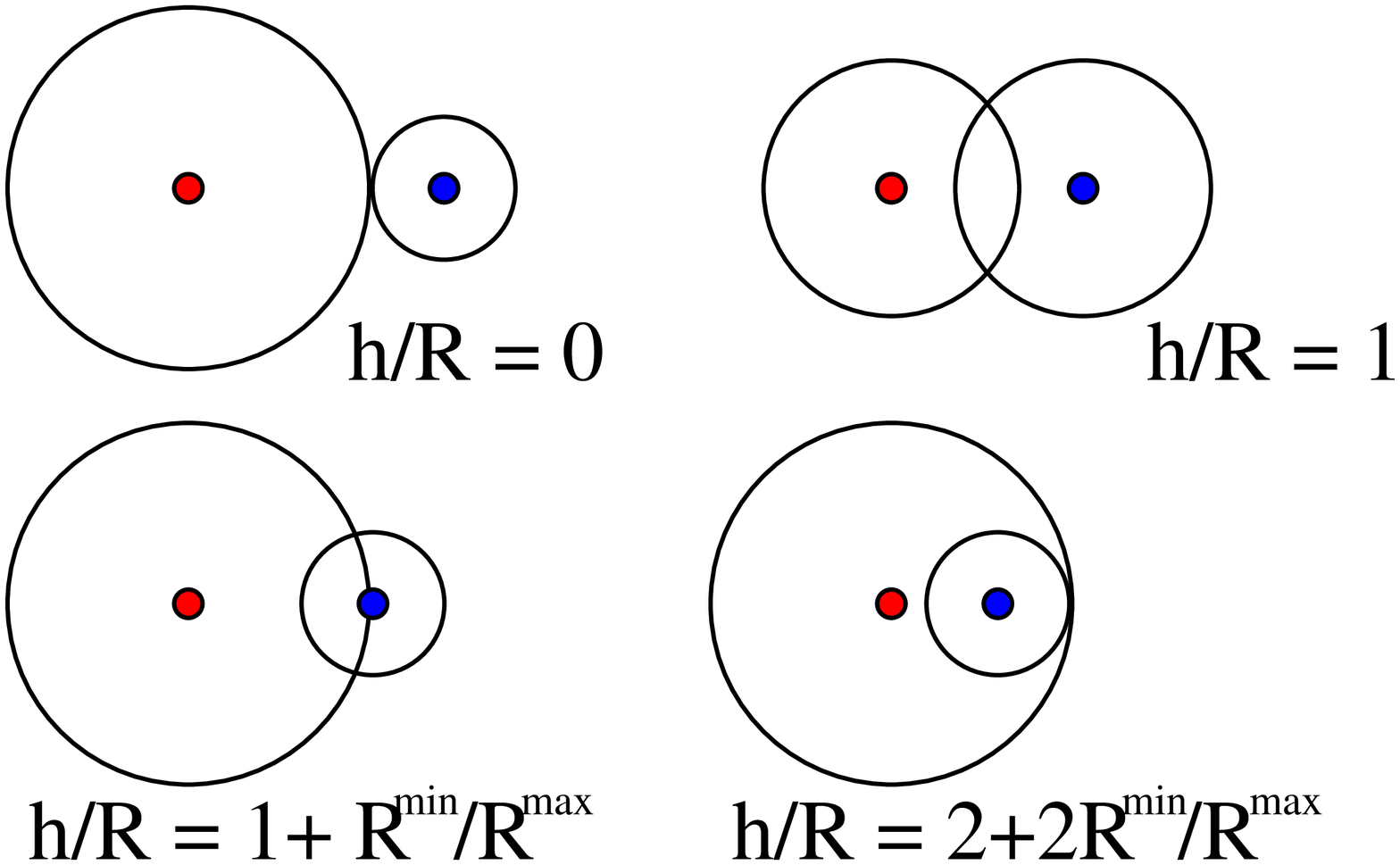}
\end{minipage}
&
\begin{minipage}{0.6\linewidth}
\includegraphics[height=6cm]{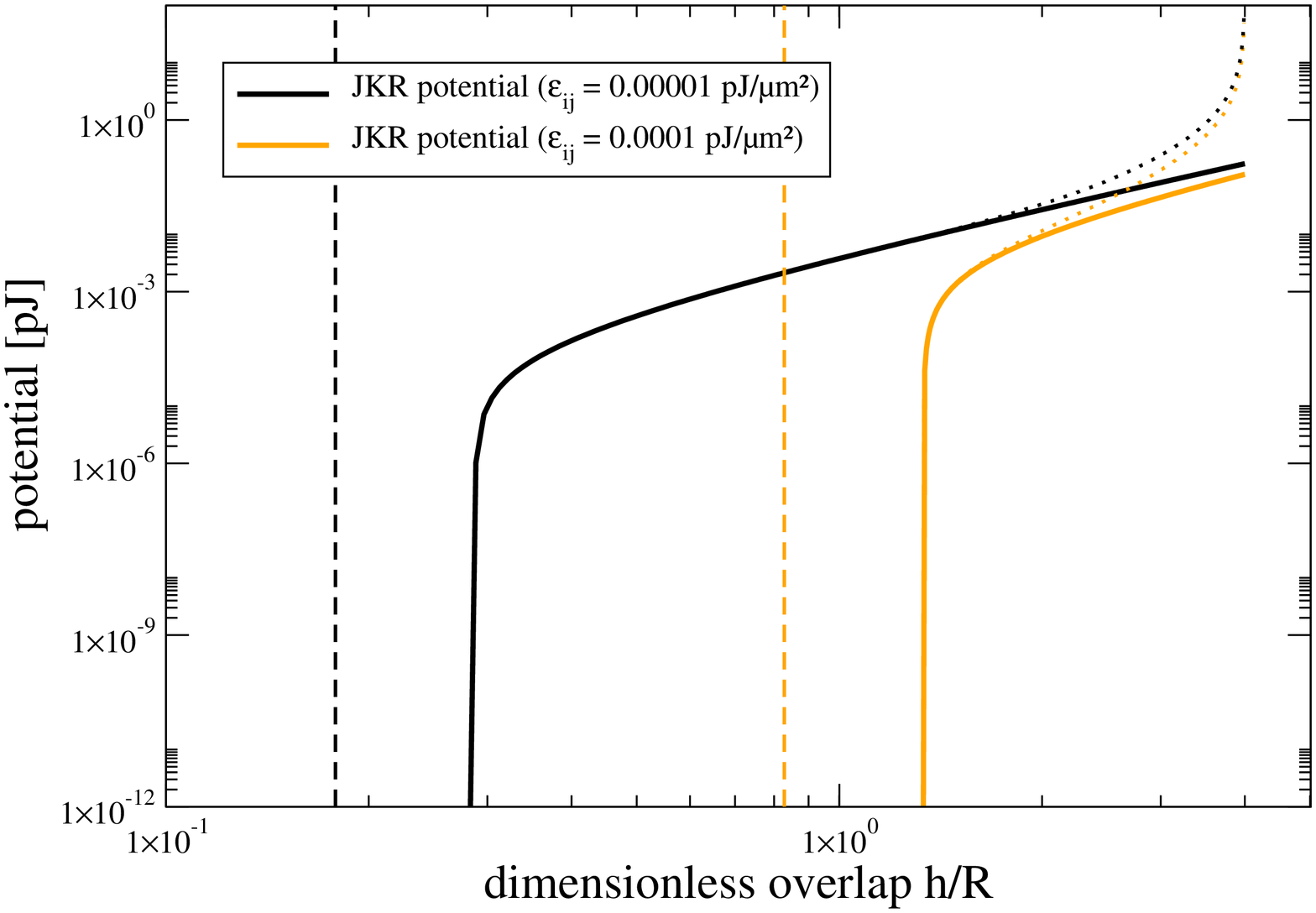}
\end{minipage}
\end{tabular}
\caption{\label{Fsphere_complete}
Sphere overlaps and modification of the interaction potential. 
{\bf Left}:
Typical contact configurations corresponding to special 
values of the dimensionless sphere distance $h_{ij}/R_{ij}$ -- for
clarity the indices have been dropped. For vanishing virtual overlap,
one obtains $h/R = 0$, whereas for the case when the centre of the
smaller sphere is placed exactly on the boundary of the
larger one, one has $h/R = 1 + R_{\rm min}/R_{\rm max}$. Complete
inclusion of the smaller sphere within the larger one 
corresponds to even larger values of $h/R$. In
this regime, JKR theory is not valid anymore. 
{\bf Right}:
For this reason, the JKR interaction has been supplemented with a pole
at complete sphere overlap. Matching has been
performed at $h_{ij}/R_{ij}=1$, where one has 
$\pi(\hat{\Vektor{r}}_i, \hat{\Vektor{r}}_j)=R_{ij}^2$. 
In the physiological regime, the potential is unchanged. Parameters have
been chosen as in figure~\ref{Fjkr_model}. 
}
\end{figure}

The quantity $h_{ij}/R_{ij}$ describes the relative position of both
spheres and is related to the orthogonal sphere distance (compare
\cite{edelsbrunner1996}) 
\bea
\pi(\hat{\Vektor{x}}, \hat{\Vektor{y}}) 
= \left(\Vektor{x} - \Vektor{y}\right)^2 - R_x^2 - R_y^2
\eea
for the spheres $\hat{\Vektor{x}} = (\Vektor{x}, R_x^2)$ and 
$\hat{\Vektor{y}} = (\Vektor{y}, R_y^2)$ via
\bea
\pi(\hat{\Vektor{r}}_i, \hat{\Vektor{r}}_j) 
= \left(\frac{h_{ij}}{R_{ij}}\right)^2 R_{ij}^2 
- 2\left(\frac{h_{ij}}{R_{ij}} - 1\right) R_i R_j\,,
\eea
compare also figure~\ref{Fsphere_complete} left panel. 

The full JKR-theory has several shortcomings:
\begin{enumerate}
\item
It is only valid for small deformations $h_{ij}/R_{ij} \ll 1$, since
the linear elastic theory assumed in the
derivation of (\ref{Ejkr_fullmodel}) is not valid for large deformations
\cite{landau1959}. In addition, it approximates the cytoskeleton as a
homogeneous solid, which is not the case \cite{verdier2003}.
Regarding the numerical solution of the interacting particle system,
this has the consequence that some cells may be completely
covered by others, since the JKR force (\ref{Ejkr_model}) does not
diverge at complete overlaps. To circumvent this, a modified
interaction potential has been used, which displays this divergence
\bea
V(x) &=& f(x) V^{\rm JKR}(x)\,,\nn\\
f(x) &=& \left\{\begin{array}{ccc}
\frac{(x_{\rm d} - x_{\rm m})^2}
{(x_{\rm d} - 2 x_{\rm m})(x_{\rm d} - x)}
-\frac{x}{x_{\rm d} - 2 x_{\rm m}} & : & x_{\rm m} \le x \le x_{\rm d}\\
1 & : & \mbox{else}
\end{array}\right.\,,
\eea
if one chooses as matching point $x_{\rm m} = 1$ and as point for
divergence $x_{\rm d} = 2 + 2 R_{\rm min}/R_{\rm max}$ 
(compare figure~\ref{Fsphere_complete}). The choice of this modified potential
only led to significantly different growth dynamics for $\order{10^4}$
cells if cellular growth was constrained by static boundaries
\cite{galle2005b}, which indicates that the used drag forces (see
appendix~\ref{ASeom}) were
small enough to enable fast relaxation. 

\item
The JKR-theory does neither include viscous effects arising from the
cytoskeleton nor dissipation occurring in the extracellular
matrix. Therefore, the model has been supplemented with additional drag
forces, which are specified in appendix~\ref{ASeom}.

\item
The original result (\ref{Ejkr_fullmodel}) has been derived as a pure
two-body interaction \cite{johnson1971}, which is also the case for
its purely elastic precursor \cite{hertz1882,landau1959}. 
However, for many adhering spheres already for small
individual deformations additional forces will come into play, since
\begin{itemize}
\item the spheres are pre-stressed,
\item the contact regions of various cells may overlap.
\end{itemize}
Thus, the JKR model does not correctly describe cellular compression
for multiple overlaps.
The extent of this shortcoming will critically depend on the current
adjacency topology which makes an analytical approach infeasible. For
numerical ease and due to missing estimates in this article the
following practitioners 
approach has been chosen. Below the target cell volume
$V_{i/j}^{\rm target}$ the cell experiences additional repulsive --
isotropic -- forces due to compression of the cytoskeleton. 
Then, the resulting additional repulsive force acts in the direction
of the neighbours $j$ with magnitude
\bea
F_{ij}^{\rm comp} &=& A_{ij}\left[
\frac{E_i}{3(1-2\nu_i)}\left(1 - \frac{V_i}{V_i^{\rm target}}\right) 
+ \frac{E_j}{3(1-2\nu_j)}\left(1 - \frac{V_j}{V_j^{\rm target}}\right)\right]\,, 
\eea
where $V_{i/j}$ denote the current cellular sphere volumes (reduced by
the overlaps with neighbouring spheres) and
$A_{ij}$ the circular JKR contact surfaces in equation
(\ref{Econtactsurface}). Note that owing to model simplicity, neither
volume nor surface corrections \cite{schaller2005a} are calculated 
with the Voronoi tessellation in this article.

\item
Whereas the forces in the approximate model (\ref{Ejkr_model}) only
depend on the actual relative cellular positions, a more realistic
scenario would have to include hysteresis effects, as adhesive
intercellular bonds form after contact \cite{chu2004}. This however
would require to keep track of the time evolution of cellular
adjacencies. 
In part, the time evolution can be incorporated into the time
dependence of the adhesive parameters
\bea
\sigma_{ij} = \frac{\sigma^{\rm max}}{2} 
\left[C_i^{\rm rec}(t) C_j^{\rm lig}(t) + C_i^{\rm lig}(t) C_j^{\rm rec}(t)\right]\,,
\eea
where the $0 \le C_{i/j}^{\rm rec/lig}(t) \le 1$ represent the
receptor or ligand
densities on the cell membrane -- normalised relative to a maximum
density, and $\sigma^{\rm max}$ is the maximum adhesion energy, respectively.
It must be noted that also the cytoskeleton reorganises and
thereby the intrinsic cell shape will not remain spherical after
contact. A full description of these effects would therefore not only
require time-dependent elastic parameters ($E_i$, $\nu_i$), but also
the implementation of a dynamically changing intrinsic 
equilibrium cell-shape, which is presumably not within the reach of a
centre-based model \cite{drasdo2003,kreft2001a}.

\item 
In addition, the derivation of the JKR model relies on the fact that
only normal forces act. For cell doublets with friction, shear forces
will in reality exist. It is assumed here that the net effect of 
shear forces on the validity of the JKR approach can be neglected,
such that they can be independently included in the drag forces.
\end{enumerate}

At least for keratinocytes the application of the JKR model to cell
doublet rupture experiments \cite{chu2004} leads to 
discrepancies between the visual equilibrium distance and the
equilibrium distance predicted by the full JKR-model (\ref{Ejkr_fullmodel}): 
If one derives via equation (\ref{Emaxadh}) the maximum adhesion energy
density from the maximum rupture force recorded in \cite{chu2004}, the
resulting equilibrium distance predicted by (\ref{Ejkr_fullmodel}) is
considerably different than observed in the figures of the same
publication: The indentation $h_{ij}$ resulting from equation
(\ref{Eambiguity}) becomes negative (pointing to extrapolation of
JKR theory beyond the region of its validity), whereas for the
approximate JKR model, the limiting condition $\sigma_{ij}/(K_{ij}
R_{ij}) \ll 1$ is certainly violated, which would lead to considerably
smaller equilibrium distances (larger indentations) than in reality.
For example, for cell-cell
contact times smaller than 30 seconds, average rupture forces of 20 nN
have been measured \cite{chu2004}. Assuming $K_{ij} = 1000$ Pa and
$R_{ij} = 2.5 \;\rm\mu m$ one would thereby find from equation
(\ref{Emaxadh}) an adhesion energy density of 
$\sigma^{\rm max}\approx 1.7 \;\rm nN/\mu m$. However, then the
equilibrium distances resulting from equations
(\ref{Ejkr_fullmodel}) or (\ref{Ejkr_model}), respectively, are inconsistent 
with the equilibrium distances in \cite{chu2004}. This indicates that
the JKR model is not directly applicable to strongly adhesive
cells. For larger times, the discrepancy becomes even worse.

However, we expect that all these shortcomings are not major sources of
error if one aims at analysing control mechanisms. An improved
contact model could generally be included in such simulations, but it
should be reasonably motivated by microscopic theories or experimental
data first. 
\section{Random Forces}\label{ASrf}


Due to thermal fluctuations, any particle in a solution will be subject
to random forces (Brownian motion). In addition, some cell types
exhibit intrinsic (active) movement which sometimes appear to be of random
nature and thus follow the same mathematics as Brownian motion. 
For systems with these characteristics, the time-dependent
stochastic forces $\Vektor{F}(t)$ modelling the random behaviour have
to fulfil two conditions \cite{ma2003}:
\begin{enumerate}
\item their mean vanishes $\expval{\Vektor{F}(t)}=0$ and
\item the forces are not correlated,
      i.~e., $\expval{\Vektor{F}(t_1) \cdot \Vektor{F}(t_2)}= 3 \xi^2
      \delta(t_1 - t_2)$.
\end{enumerate}
The parameter $\xi$ thereby quantifies the strength of the
stochastic fluctuations.
The movement of single cells in a solution is highly overdamped
\cite{dallon2004}, and any stochastic force fulfilling the above conditions
will lead in the Langevin equation to a diffusion-like evolution of
cellular distribution, i.~e., in the 
absence of additional forces the squared displacement will be given by
\bea
\expval{\left[\Vektor{r}(t) - \Vektor{r}_0\right]^2} = 3\frac{\xi^2}{\gamma^2}
t = 6 D t\,,
\eea
where $D$ is the corresponding diffusion coefficient and
$\gamma$ is a dampening constant, which effectively describes the strength of
friction. The above identity is also known as the fluctuation
dissipation theorem, since it connects the fluctuations ($\xi$)
with the dissipation ($\gamma$). If this dynamics is observed for free
spherical cells in medium, the friction constant
can for highly-damped dynamics be well approximated by the Stokes
friction $\gamma_0 = 6 \pi \eta R$,
where $R$ represents the radius of the cell and
$\eta$ the viscosity of the surrounding medium. Evidently, with the
same random forces applied, cellular movement will be much smaller if
drag forces due to cellular bindings are at work. 
For numerical implementations, a stochastic force
fulfilling the above conditions can be given by \cite{ma2003}
\bea\label{Erandom}
F_i(t) &=& \frac{\xi}{\sqrt{\Delta t}} \,\chi^{\rm GAUSS}_{0, 1}\,,
\eea
where $\Delta t$ describes the width of the timestep, and $\chi^{\rm
GAUSS}_{0, 1}$ is a random number drawn from a Gaussian distribution \cite{matpack_manual}
with mean $\mu_{\chi}=0$ and width $\sigma_{\chi}=1$. 

It should be noted however, that active random eucaryotic movement
in reality usually occurs with pseudopods \cite{fletcher2004}: The cell attaches
protrusions to neighbouring cells (or the extracellular matrix) and
randomly pulls towards them. This has two further implications
\begin{itemize}
\item
the stochastic forces become two-body forces, i.~e., the neighbour cell that the
pseudopod is attached to, is subject to the corresponding negative
force.
Also, the forces act into the direction of the normals \cite{galle2005a}.
\item
Since the pseudopods do not enable pushing, the average stochastic
force component into the direction of a given neighbour cell will not
in general vanish. For example, at interfaces of dense 
tissue (where the pseudopods find resistance) and fluid (where no net
force can be generated) one cannot expect the contributions into the
different directions to compensate each other.
\end{itemize}

Since the intrinsic logic behind active cellular movement following
Brownian mathematics is not fully understood and also active movement
with pseudopods is not quantified for the cell types considered in the
simulation (keratinocytes and melanocytes), we have chosen to
implement stochastic forces via equation (\ref{Erandom}) as acting
randomly on every cell that reacts passively to these in return.

\section{Equations of Motion}\label{ASeom}

For $N$ spherical cells with positions $\Vektor{x}_i(t)$ and radii $R_i$
subject to cell-cell as well as cell-medium and
cell-substrate interactions, the equations of motion can in the
reference frame of motionless medium and boundaries be summarised
as (compare also \cite{ferrez2001})
\bea\label{Efullmodel}
m_i \ddot{x}_i^\alpha &=& 
F_i^\alpha + \sum_{J \in {\cal NB}(i)} F_{iJ}^\alpha
+ \sum_{j \in {\cal NN}(i)} F_{ij}^\alpha
- \sum_\beta \gamma_i^{\alpha\beta} \dot{x}_i^\beta\nn\\
&&- \sum_{J \in {\cal NB}(i)} \sum_\beta \gamma_{iJ}^{\alpha\beta}
\left[\dot{x}_i^\beta + R_i \left(\Vektor{n}_{iJ} \times
\Vektor{\omega}_i\right)^\beta \right]
- \sum_\beta \sum_{j \in {\cal NN}(i)} \gamma_{ij}^{\alpha\beta} 
\left[\dot{x}_i^\beta - \dot{x}_j^\beta 
+ R_i \left(\Vektor{n}_{ij}\times\Vektor{\omega}_i\right)^\beta
+ R_j \left(\Vektor{n}_{ij}\times\Vektor{\omega}_j\right)^\beta\right]
\nn\,,\\
I_i \dot{\omega}_i^\alpha &=&
T_i^\alpha 
+ \sum_{J \in {\cal NB}(i)} R_i \left(\Vektor{n}_{iJ} \times
\Vektor{F}_{iJ}\right)^\alpha
+ \sum_{j \in {\cal NN}(i)} R_i \left(\Vektor{n}_{ij} \times
\Vektor{F}_{ij}\right)^\alpha
- \sum_\beta \Gamma_i^{\alpha\beta} \omega_i^\beta\nn\\
&&- \sum_{J \in {\cal NB}(i)} \sum_\beta \Gamma_{iJ}^{\alpha\beta}
\left[\dot{x}_i^\beta + R_i \left(\Vektor{n}_{iJ} \times
\Vektor{\omega}_i\right)^\beta \right]
- \sum_\beta \sum_{j \in {\cal NN}(i)} \Gamma_{ij}^{\alpha\beta} 
\left[\dot{x}_i^\beta - \dot{x}_j^\beta 
+ R_i \left(\Vektor{n}_{ij}\times\Vektor{\omega}_i\right)^\beta
+ R_j \left(\Vektor{n}_{ij}\times\Vektor{\omega}_j\right)^\beta\right]\,,\nn\\
\eea
where $\alpha, \beta \in \{0,1,2\}$ denote the Cartesian indices and 
$i,j \in \{0,1,\ldots,N-1\}$ the cellular indices. The first equation
describes the evolution of the cell positions $\Vektor{x}_i(t)$, whereas
the second equation accounts for the evolution of the cellular spin
velocities $\Vektor{\omega}_i(t)$. The terming ${\cal NN}(i)$ denotes all
cells having direct contact with cell $i$, whereas ${\cal NB}(i)$
refers to all boundaries in direct contact with cell $i$. Such a
set of neighbouring cells can be efficiently determined as a subset of all
neighbours in the weighted Delaunay triangulation of the set of
spheres. (We had developed and applied such a triangulation module
previously in \cite{schaller2004,schaller2005a}.) Since
for most problems few boundary conditions will be given, these are
hard-wired in the code for every specific problem individually. Note
that the back-reaction of the cells on the boundaries is
neglected implicitly assuming that the boundaries are stationary.

The first term on the right-hand side of the
first equation $F_i^\alpha$ may generally 
include deterministic (for example, crawling forces on a substrate)
and stochastic (e.~g.~Brownian motion) forces on a single cell,
whereas the second and third terms
$\sum_{j \in {\cal NB}(i)} F_{iJ}^\alpha$ and
$\sum_{j \in {\cal NN}(i)} F_{ij}^\alpha$ include the cell-boundary
and intercellular two-body forces (e.~g.~stochastic two-body forces or
the deterministic JKR-force, compare subsections~\ref{ASrf}
and~\ref{ASjkr}), respectively. The fourth term 
$\sum_\beta \gamma_i^{\alpha\beta} \dot{x}_i^\beta$
denotes cell-medium friction, whereas the last two terms denote
dampening due to friction with the boundaries
($\gamma_{iJ}^{\alpha\beta}$) and with neighbouring cells
($\gamma_{ij}^{\alpha\beta}$), respectively. Note that the dampening
forces can be divided in a contribution proportional to a relative
cell velocity and a contribution arising from the angular velocities
of both cells, where the normal vector $\Vektor{n}_{ij}$ is understood to
point from cell $i$ towards cell $j$ (which restores the
apparently violated antisymmetry of the dampening forces under
exchange of $i$ and $j$).

The quantity $I_i$ on the left-hand side of the second equation
denotes the inertial momentum ($I_i = \frac{2}{5} m_i R_i^2$ for rigid
homogeneous spheres). 
In analogy with the forces, the first term on the right-hand side of
the second equation  $T_i^\alpha$ describes an intrinsic torque
of the $i^{\rm th}$ cell, whereas the second and third terms describe
the torques generated by cell-boundary and cell-cell interactions. 
In contrast to the forces, the dampening constant
$\Gamma_i^{\alpha\beta}$ (capital coefficients denote
rotational dampening) does not only incorporate friction with the
surrounding fluid but also the dissipation of rotational energy into
internal degrees of freedom of the cell (i.~e., finally heat).
As with the forces, the last terms describe the rotational dampening
due to cell-boundary and cell-cell interaction, respectively.

Note the equation for the rotation may generally back-react
onto the first equation via other channels as well. For example, the
interaction forces could depend on respective angular
momenta. 
Moreover, the terms describing the influence of the torques on the
angular velocity implicitly assume that the cell is a rigid body,
which is not the case.
Although already sophisticated enough, the above equations
should therefore be regarded a simple possible ansatz.

In the over-damped approximation
\bea
m_i\ddot{x}_i^\alpha \approx 0 \qquad \mbox{and}\qquad
I_i \dot{\omega}_i^\alpha \approx 0\,,
\eea
which is widely used to describe cell movements in fluids, the
interaction forces and torques are always balanced by the friction
forces and torques, respectively.
Concerning the friction torques we assume that
the internal friction of the cell is dominant 
($\Gamma_i^{\alpha\beta} \gg \Gamma_{ij}^{\alpha\beta},
\Gamma_{iJ}^{\alpha\beta}$) and that there is no intrinsic torque
generated by the cell types we consider ($T_i^\alpha=0$).
In addition, we assume that the dominant torque dampening is
approximately isotropic
($\Gamma_i^{\alpha\beta}=\delta^{\alpha\beta}\Gamma_i$).
Then, one obtains for the angular velocity
\bea\label{Ecellspin}
\Gamma_i \omega_i^\alpha \approx 
 \sum_{J \in {\cal NB}(i)} R_i \left(\Vektor{n}_{iJ} \times
\Vektor{F}_{iJ}\right)^\alpha
+ \sum_{j \in {\cal NN}(i)} R_i \left(\Vektor{n}_{ij} \times
\Vektor{F}_{ij}\right)^\alpha\,.
\eea
If one inserts the above expression into the first equation of
(\ref{Efullmodel}), one observes that the friction terms describing
the influence of the torque on the cellular force dampening is
suppressed by prefactors of 
$\gamma_{iJ}^{\alpha\beta} R_i^2/\Gamma_i$ and 
$\gamma_{ij}^{\alpha\beta} R_i^2/\Gamma_i$. Whether these terms can be
neglected, is dominantly related to the structure of the
cytoskeleton. We assume here that the cytoskeleton does not transmit
shear forces well.

With these approximations, the relevant equations of motion take the form 
\bea
0 &\approx& 
F_i^\alpha + \sum_{J \in {\cal NB}(i)} F_{iJ}^\alpha
+ \sum_{j \in {\cal NN}(i)} F_{ij}^\alpha
- \sum_\beta \gamma_i^{\alpha\beta} \dot{x}_i^\beta
- \sum_{J \in {\cal NB}(i)} \sum_\beta \gamma_{iJ}^{\alpha\beta}
\dot{x}_i^\beta
- \sum_\beta \sum_{j \in {\cal NN}(i)} \gamma_{ij}^{\alpha\beta} 
\left(\dot{x}_i^\beta - \dot{x}_j^\beta\right)\,.
\eea
Note that solving the remaining equation for the cellular spin
velocities (\ref{Ecellspin}) is not necessary, since its back-reaction
on the cell movement has been neglected and any snapshot of a rotating
sphere cannot be distinguished from a motionless sphere.
The above equation can be rewritten to yield
\bea\label{Eeom}
\sum_{k, \beta} \left\{ \left[
\gamma_k^{\alpha\beta} 
+ \sum_j \gamma_{kj}^{\alpha\beta}
+ \sum_J \gamma_{kJ}^{\alpha\beta}\right] \delta_{ik} - \gamma_{ik}^{\alpha\beta}\right\}
	\dot{x}_k^\beta = F_i^\alpha + \sum_j F_{ij}^\alpha\,.
\eea
From the properties of the friction coefficients it can also be
deduced that the linear system defined by this equation is
symmetric and also diagonally dominant as long as 
$\gamma_i^{\alpha\alpha} + \sum_J \gamma_{iJ}^{\alpha\alpha} > 0
\qquad \forall i, \alpha$.
In addition, it must be noted that the system will be extremely
sparsely populated as the friction coefficients  
vanish for all cells not being in direct contact, see
the appendix~\ref{ASns} for an example. 

A usual choice for cell-medium friction is the well-known
Stokes-relation $\gamma_i^{\alpha\beta} = 6 \pi \eta R_i
\delta^{\alpha\beta}$ \cite{joos1989} introduced in
subsection~\ref{ASrf}.  
The friction coefficients and two-body forces fulfil the following
conditions:
\bea\label{Esymmetry_conditions}
\gamma_{ij}^{\alpha\beta} = \gamma_{ji}^{\alpha\beta}\,, 
\qquad
F_{ij}^\alpha &=& -F_{ji}^\alpha
\qquad\qquad\mbox{(Newton's third axiom)}\,,\nn\\
\gamma_{ij}^{\alpha\beta} &=& \gamma_{ij}^{\beta\alpha} 
\qquad\qquad\;\,\mbox{(projection operator property)}\,,\nn\\
\gamma_i^{\alpha\beta} &=& \gamma_i \delta^{\beta\alpha}
\qquad\;\;\;\;\;\;\mbox{(isotropy)}\,,\nn\\
\gamma_{ii}^{\alpha\beta} &=& 0 
\qquad\qquad\;\;\;\;\;\,\mbox{(no self-friction)}\,.
\eea
In a strict sense, Newton's third axiom only applies to the total
two-body force. However, here the model should consistently include
contact forces $F_{ij}^\alpha$ and drag forces 
$\gamma_{ij}^{\alpha\beta}\left(\dot{x}_i^\beta-\dot{x}_j^\beta\right)$,
which may act independently from each other. Therefore,
{\em actio et reactio} has been assumed to act separately. The
symmetry in $(\alpha,\beta)$ of the friction coefficients also arises from
the symmetry properties of the projection operators:
The drag forces expressed by the friction coefficients
may be divided in normal drag forces and tangential (shear) drag forces.
Assuming that they are proportional to the effective contact area
between two cells $i$ and $j$
\bea
A_{ij}^{\rm eff} = A_{ij} \frac{1}{2}\left[
	C_i^{\rm rec}(t) C_j^{\rm lig}(t) + C_i^{\rm lig}(t) C_j^{\rm rec}(t)
	\right]
\eea
and to the normal or tangential projection of the velocity
differences, respectively, the friction coefficients
take the form
\bea\label{Efriction_symmetry}
\gamma_{ij}^{\alpha\beta} &=& A_{ij}^{\rm eff}\left(
	\gamma_\parallel {\cal P}_{ij, \parallel}^{\alpha\beta}
	+ \gamma_\perp {\cal P}_{ij, \perp}^{\alpha\beta}
	\right)\left(1 - \delta_{ij}\right)\,,\nn\\
\gamma_{iJ}^{\alpha\beta} &=& A_{iJ}
\left(\gamma_\parallel {\cal P}_{iJ, \parallel}^{\alpha\beta}
+ \gamma_\perp {\cal P}_{iJ, \perp}^{\alpha\beta}
\right)\,.
\eea
The friction constant $\gamma_\perp$ predominantly describes
internal friction within the cytoskeleton \cite{galle2005a}, since
force contributions for movements normal to the cell-cell
contact surface are already contained within the JKR interaction
model. The friction constant $\gamma_\perp$ is
set to vanish within this article thereby 
implicitly assuming that dampening due to friction within the
cytoskeleton is much smaller than dampening due to cell-cell
bindings. In contrast, the tangential 
friction constant $\gamma_\parallel$ describes drag forces resulting
from broken bindings during movements tangential to the intercellular
contact plane \cite{dallon2004,galle2005a}. 
For model consistency, the used contact surfaces are chosen identical
with the JKR contact surface (\ref{Econtactsurface}). Since
over a wide range of physiological overlaps this relates to the
spherical overlap that would result from undeformed spheres by about a
factor of two, a different choice of the contact surface could be
compensated by appropriately changed friction parameters.

The intercellular tangential and perpendicular projectors are given by
\bea\label{Eproj_cc}
{\cal P}_{ij, \parallel}^{\alpha\beta} =
\delta^{\alpha\beta} - n_{ij}^\alpha n_{ij}^\beta\,, \qquad 
{\cal P}_{ij, \perp}^{\alpha\beta} =
n_{ij}^\alpha n_{ij}^\beta\,,
\eea
and the cell-boundary projectors
\bea\label{Eproj_cb}
{\cal P}_{iJ, \parallel}^{\alpha\beta} =
\delta^{\alpha\beta} - n_{iJ}^\alpha n_{iJ}^\beta\,, \qquad 
{\cal P}_{iJ, \perp}^{\alpha\beta} =
n_{iJ}^\alpha n_{iJ}^\beta\,,
\eea
respectively. 
In the above projection operators, $\Vektor{n}_{ij}$
represents the normal vector pointing from cell $i$ towards cell $j$
(compare also figure~\ref{Ffriction_example} left panel), whereas
$\Vektor{n}_{iJ}$ denotes the normal vector of the boundary $J$ at the
contact point with cell $i$. 
Note that with these projection operators, the conditions on the
friction coefficients (\ref{Esymmetry_conditions}) are automatically
fulfilled.

\section{Numerical Solution}\label{ASns}

\begin{figure}[htb]
\includegraphics[height=6cm]{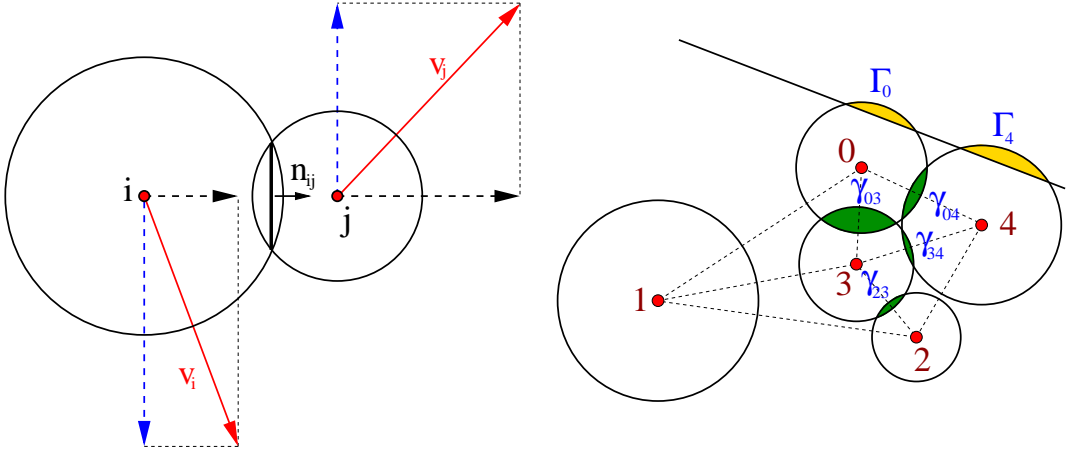}
\caption{\label{Ffriction_example}
Two-dimensional illustration of intercellular force calculation.
{\bf Left}: The differential velocities between two cells $i$ and $j$
(circles) in contact may give rise to different drag forces, one proportional to the
tangential part of the velocity difference and one proportional to the
perpendicular part as indicated with dashed lines.
{\bf Right}:
The dotted lines denote the weighted Delaunay triangulation of the set of
spheres. Only spheres with cell-cell contact (overlap, dark grey)
will contribute to intercellular friction and to intercellular forces. 
Cell-boundary contact (overlap, light grey) will contribute to
cell-medium friction and to cell-boundary forces.
Spheres that are not connected in the weighted Delaunay
triangulation will not interact.
}
\end{figure}

An example including cell-cell and cell-boundary friction is
illustrated in figure~\ref{Ffriction_example}. 
Indeed, for this special example all non-isotropic friction
coefficients vanish except
$\gamma_{03}^{\alpha\beta}, \gamma_{04}^{\alpha\beta}, 
\gamma_{23}^{\alpha\beta}, \gamma_{34}^{\alpha\beta}, 
\Gamma_0^{\alpha\beta}, \Gamma_4^{\alpha\beta}$. 
Consequently, for this example the system (\ref{Eeom}) 
would assume the form
\bea
\left(
\begin{array}{ccccc}
\f{\Gamma}_0 + \f{\gamma}_0 + \f{\gamma}_{03} + \f{\gamma}_{04} & 
{\mathbb O} &
{\mathbb O} &
-\f{\gamma}_{03} &
 -\f{\gamma}_{04}\\
{\mathbb O} &
 \f{\gamma}_1 & 
{\mathbb O} &
{\mathbb O} &
{\mathbb O}\\
{\mathbb O} &
{\mathbb O} &
 \f{\gamma}_2 + \f{\gamma}_{23} & 
 -\f{\gamma}_{23}  &  
{\mathbb O}\\
 -\f{\gamma}_{03} &  
{\mathbb O} &
 -\f{\gamma}_{23} &  
 \f{\gamma}_3 + \f{\gamma}_{03} + \f{\gamma}_{34} &  
 -\f{\gamma}_{34}\\
 -\f{\gamma}_{04} &  
{\mathbb O} &
{\mathbb O} &
 -\f{\gamma}_{34} &  
 \f{\gamma}_4 + \f{\gamma}_4 + \f{\gamma}_{04} + \f{\gamma}_{34}
\end{array}
\right)
\left(
\begin{array}{l}
\dot{\Vektor{x}}_0\\
\dot{\Vektor{x}}_1\\
\dot{\Vektor{x}}_2\\
\dot{\Vektor{x}}_3\\
\dot{\Vektor{x}}_4
\end{array}
\right)
=\nn\\
\left(
\begin{array}{l}
\Vektor{F}_0 + \Vektor{F}_{03} + \Vektor{F}_{04} + \Vektor{F}_{0\rm B}\\
\Vektor{F}_1\\
\Vektor{F}_2 + \Vektor{F}_{23}\\
\Vektor{F}_3 + \Vektor{F}_{30} + \Vektor{F}_{32} + \Vektor{F}_{34}\\
\Vektor{F}_4 + \Vektor{F}_{40} + \Vektor{F}_{42} + \Vektor{F}_{43} +
\Vektor{F}_{4\rm B}\\
\end{array}
\right)\,,\nn\\
\eea
where in three dimensions the symbols $\Vektor{F}_i$ and $\Vektor{x}_i$
denote vectors in ${\mathbb R}^3$ and ${\mathbb O}$, 
$\f{\gamma}_{ij}$, $\f{\gamma}_i$, and $\f{\Gamma}_{i}$ denote $3 \times 3$
matrices.
This system is evidently symmetric, sparsely populated and weakly diagonally
dominated, since \mbox{$\gamma_i^{\alpha\alpha} + \Gamma_i^{\alpha\beta} > 0
\qquad \forall \qquad i,\alpha$}. In addition, all friction
coefficients are positive. Gershgorin's circle theorem then suffices to
guarantee positive definiteness of the dampening matrix. The
number of next neighbours in contact corresponds to the number of
off-diagonal blocks in the dampening matrix, such that the system
becomes extremely sparse for large matrices.

Such systems can for efficiency be supplemented with the weighted
Delaunay triangulation of a set of spheres for adjacency detection
\cite{schaller2004}. Since the dampening matrix is positive definite,
the method of conjugate gradients \cite{shewchuk1994} is well suited
to the problem. However, since realistic systems will contain much more
than 5 cells, the matrices would not fit into main memory, if stored
completely. Fortunately, the matrices are only sparsely populated and
the method of conjugate gradients can efficiently be combined with a 
row-indexed sparse storage scheme \cite{press1994} to compute a solution 
$\dot{x}_i^\alpha$. Note that the solution of the full system is an
improvement over existing models: For example, in \cite{dallon2004} the
tangential projector ${\cal P}_{ij}^{\alpha\beta}$ had for simplicity
been approximated with the identity operator and in
\cite{schaller2005a}, the system was assumed to be diagonal.

The reaction-diffusion equation for the molecules
\bea
\pdiff{u}{t} = \nabla \left[ 
D(\Vektor{r}; t) \nabla u \right] + Q(\Vektor{r}; t)
\eea
could in principle be solved using the method of Green functions along 
the lines of \cite{newman2004}. 
However, since spatiotemporally varying diffusion coefficients as well as 
nontrivial boundary conditions are considered here, the implementation of this method 
would require an enormous amount of work.
Therefore, we have discretized above equation using the discrete 
element method on a cubic lattice.
The arising coupled system of ordinary differential equations was then solved
using the Crank-Nicolson scheme \cite{gershenfeld2000} and the 
algorithm of biconjugate gradients \cite{press1994}. 
Note that due to the strongly
varying diffusion coefficients in cornified and non-cornified
tissue (see table~\ref{Tparameters}), the steady-state approximation
is not applicable. The lattice constants for the rectangular
reaction-diffusion grid discretisation have been chosen 
larger than the cellular diameters to ensure for validity of the
discretisation approximation \cite{schaller2005a}. Note that the
reaction diffusion equation for the nutrients is made positive
definite by cells entering necrosis below critical nutrient
concentrations -- necrotic cells do not consume 
nutrients.
While the timestep of the simulation is determined by a
maximum spatial cellular stepsize (fixed at $0.5 \;\rm\mu m$), the timestep of the
reaction-diffusion grids has been divided into several substeps such
that the Courant factor \cite{press1994} is smaller than 1 in order to
increase numerical accuracy. 
To connect the discrete reaction-diffusion grid
with the spatially-continuous cellular positions, tri-linear
interpolation has been used. In addition, the concentration has been
fixed to vanish at grid nodes whose elementary cell did not contain cells. At
grid nodes in the vicinity of cells the diffusive 
properties have been smoothly adapted in the range of the values given
in table~\ref{Tparameters}.



\end{document}